\documentclass[fleqn,usenatbib]{mnras}

\usepackage{newtxtext,newtxmath}
\usepackage{array}

\usepackage[T1]{fontenc}

\DeclareRobustCommand{\VAN}[3]{#2}
\let\VANthebibliography\thebibliography
\def\thebibliography{\DeclareRobustCommand{\VAN}[3]{##3}\VANthebibliography}

\usepackage{graphicx}	% Including figure files
\usepackage{amsmath}	% Advanced maths commands
\newcommand\cm{{\rm\thinspace cm}}

\newcommand\erg{{\rm\thinspace erg}}

\newcommand\K{{\rm\thinspace K}}
\newcommand\keV{{\rm\thinspace keV}}
\newcommand\km{{\rm\thinspace km}}
\newcommand\kpc{{\rm\thinspace kpc}}

\newcommand\Msun{\hbox{$\rm\thinspace M_{\odot}$}}

\newcommand\s{{\rm\thinspace s}}
\newcommand\yr{{\rm\thinspace yr}}

\newcommand\cmsq{\hbox{$\cm^2\,$}}
\newcommand\cmcu{\hbox{$\cm^3\,$}}

\newcommand\ergps{\hbox{$\erg\s^{-1}\,$}}

\newcommand\kmps{\hbox{$\km\s^{-1}\,$}}

\newcommand\Msunpyr{\hbox{$\Msun\yr^{-1}\,$}}

\newcommand\pcmsq{\hbox{$\cm^{-2}\,$}}

\newcommand\ps{\hbox{$\s^{-1}\,$}}
\newcommand\psqcm{\hbox{$\cm^{-2}\,$}}

\title[Hidden Cooling Flows II]{Hidden Cooling Flows in  Clusters of Galaxies II: A Wider Sample}

\author[A. C. Fabian et al.]{
A. C. Fabian,$^{1}$\thanks{E-mail: acf@ast.cam.ac.uk },  J.S. Sanders$^{2}$, G.J. Ferland$^{3}$, B.R. McNamara$^{4}$, C. Pinto$^{5}$ and S.A. Walker$^6$
\\
$^{1}$Institute of Astronomy, University of Cambridge, Madingley Road, Cambridge CB3 0HA, UK\\
$^{2} $Max-Planck-Institut fur extraterrestrische Physik, Giessenbachstrasse 1, 85748 Garching, Germany\\
$^{3} $Department of Physics, University of Kentucky, Lexington KY 40506, USA\\
$^{4} $Department of Physics and Astronomy, University of Waterloo, 200 University Avenue West, Waterloo, ON N2L 3G1, Canada\\
$^{5} $INAF-IASF Palermo, Via U. La Malfa 153, I-90146 Palermo, Italy\\
$^6$ Department of Physics and Astronomy, The University of Alabama in Huntsville, Huntsville, AL 35899, USA\\ }
\date{Accepted XXX. Received YYY; in original form ZZZ}

\pubyear{2022}

\begin{document}
\label{firstpage}
\pagerange{\pageref{firstpage}--\pageref{lastpage}}
\maketitle

\begin{abstract}
We have recently uncovered Hidden Cooling Flows (HCFs) in the XMM RGS spectra of 3 clusters of galaxies, Centaurus, Perseus and A1835.  
Here we search for them in a wider sample of objects: the X-ray brightest group NGC5044; 4 moderate X-ray luminosity clusters Sersic 159, A262, A2052 and RXJ0821; and 3 high X-ray luminosity clusters RXJ1532, MACS 1931 and the Phoenix cluster. Finally we examine two Virgo elliptical galaxies, M49 and M84. All statistically allow the addition of an HCF. We find  a significant detection of an HCF in 6 clusters and 2 elliptical galaxies.  The hidden mass cooling rates are $5-40\Msunpyr$ for the normal clusters, $1000\Msunpyr$ or more for the extreme clusters and $1-2\Msunpyr$ for the elliptical galaxies. We discuss the implications of the results for the composition of the innermost parts of the massive host galaxies and look forward to future observations.
\end{abstract}


%%%%%%%%%%%%%%%%%%%%%%%%%%%%%%%%%%%%%%%%%%%%%%%%%%

%%%%%%%%%%%%%%%%% BODY OF PAPER %%%%%%%%%%%%%%%%%%

\section{Introduction}
The most massive galaxies tend to be Brightest Cluster Galaxies (BCG) dominating the mass at the centre of the deepest galaxy-scale gravitational potential wells in the Universe. 
The temperature of the  hot intracluster gas surrounding at least half of them drops inward from the cluster (or group) virial temperature $T_{\rm V}$ of tens of millions K to form a cool core. The radiative cooling time of the gas falls well below the age of the cluster within the inner 20kpc so that a cooling flow of gas into the centre can occur. Feedback of energy from a central active nucleus can reduce the level of cooling. The pertinent question remains of just how strong is the reduction. In particular, how much gas cools below $10^7\K$?

We have started to re-examine this question \citep{Fabian22}, hereafter HCF1, using X-ray spectra of cluster cores taken with the XMM Reflection Grating Spectrometer (RGS). Earlier work with RGS spectra showed little evidence of gas below about  $T_{\rm V}/3$ \citep{Peterson2001, Kaastra2001} . Strong reduction  of full cooling flows was indicated  with mass cooling rates reduced to less than 10 per cent of what was expected with no suppression \citep{Peterson2003, Peterson2006, Liu2019, Liu2021}. 
It was as if gas is cooling but only down to temperatures of 1 or 2 keV. Significant detections of some cooler gas were obtained later from emission-line indicators such as OVII and FeXVII \citep{Sanders2011, Pinto2016}.

\begin{figure}
    \centering
    \includegraphics[width=0.48\textwidth]{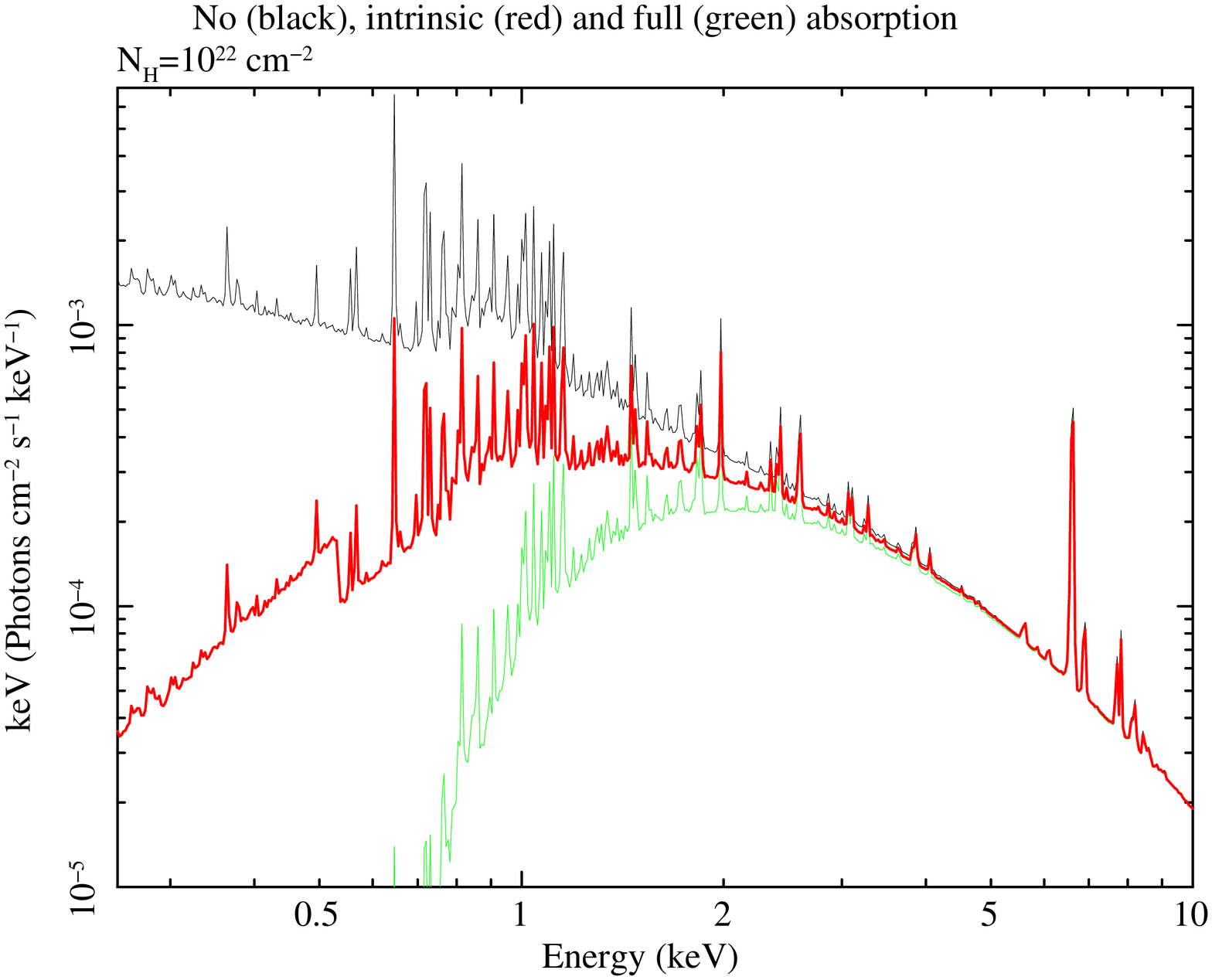}
      \includegraphics[width=0.48\textwidth]{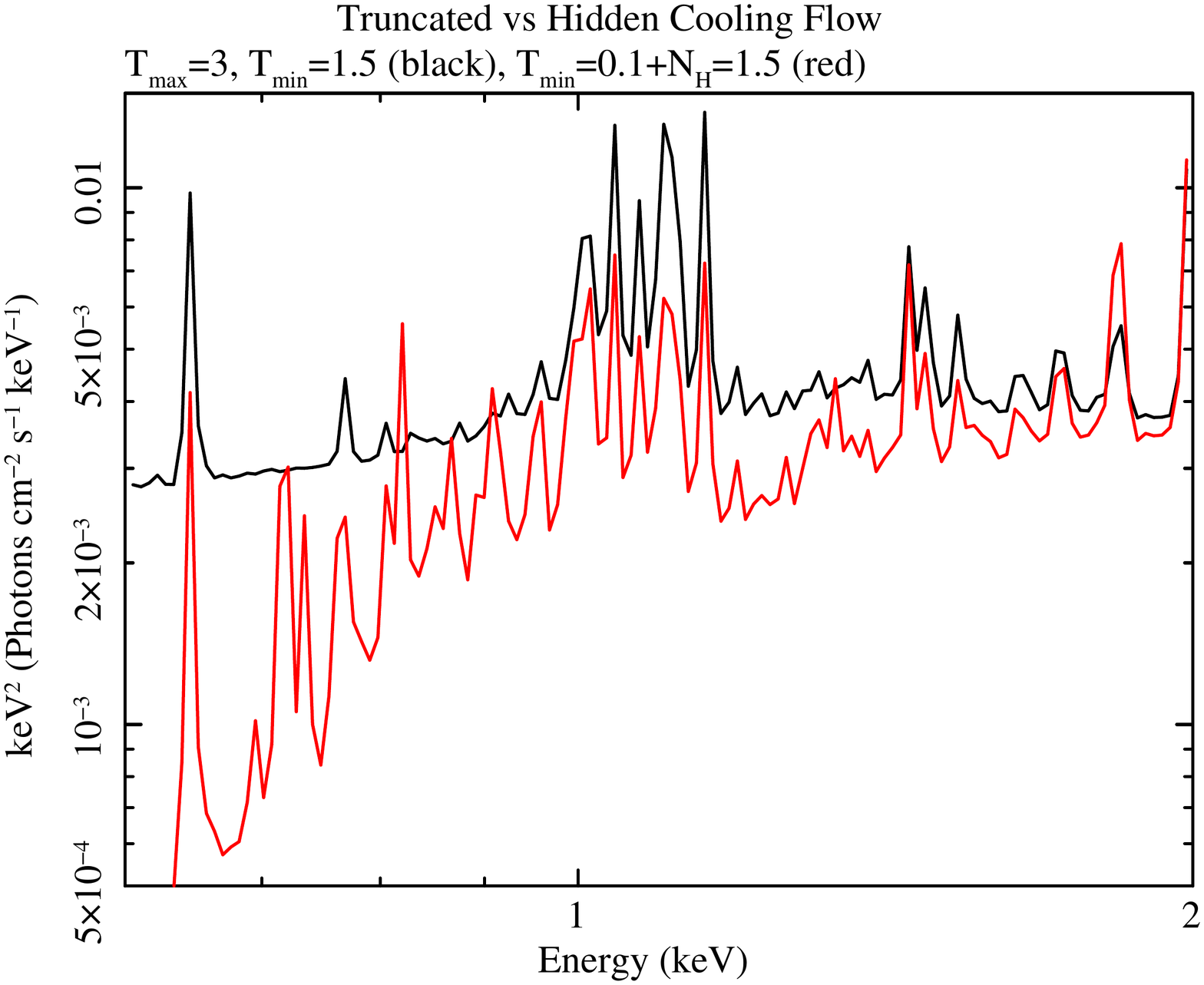} 
      \includegraphics[width=0.48\textwidth]{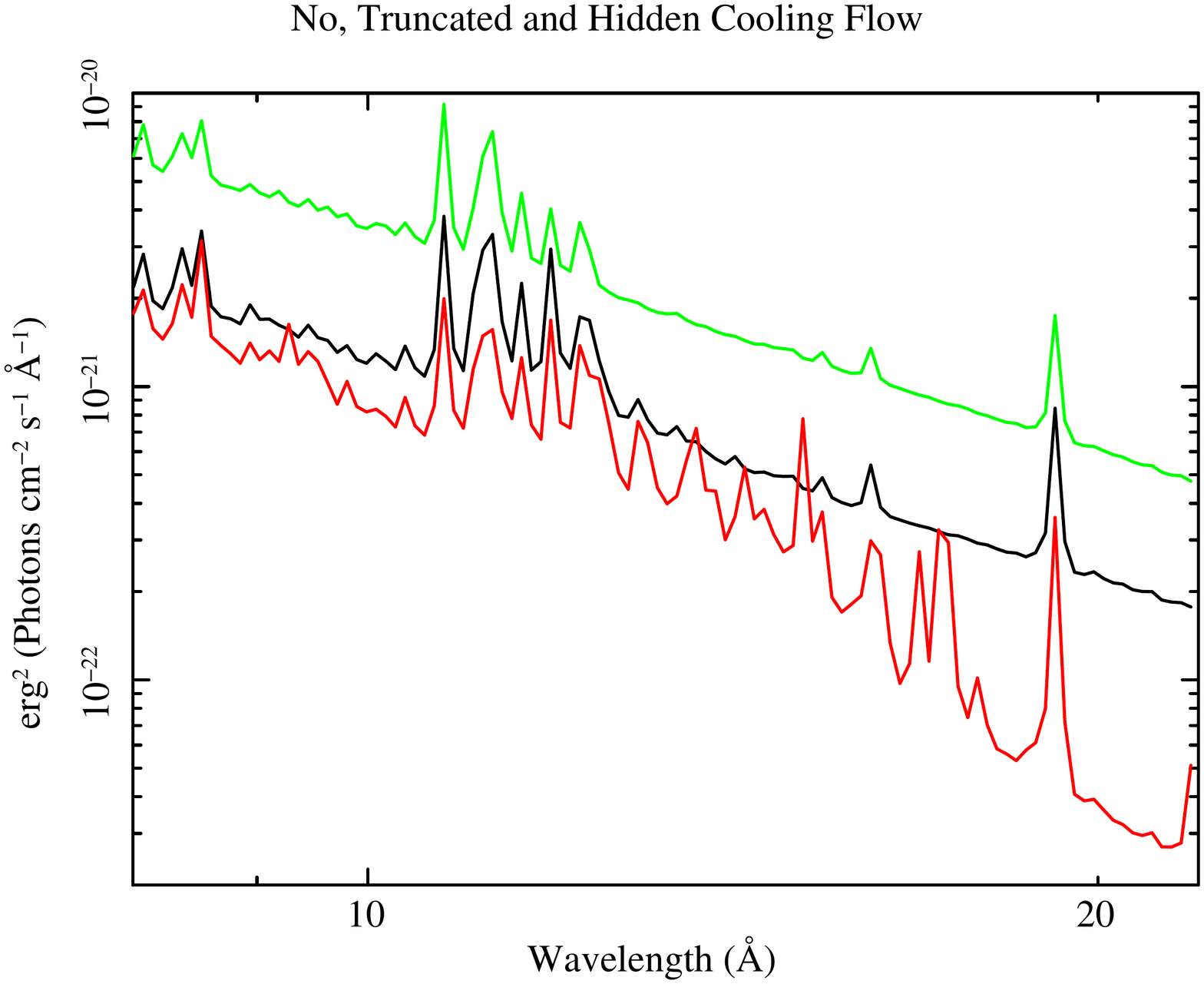}
    \caption{ Examples of model spectra including intrinsic absorption. The upper panel shows a cooling flow from a temperature of 5 keV without any absorption (top, black); with intrinsic absorption of total column density $10^{22}\pcmsq$ (middle, red); and full absorption of $10^{22}\pcmsq$ (bottom, green).  The middle panel shows a cooling flow from 3 keV truncated at 1.5 keV (top, black) and cooling from 3 keV to 0.1 keV with intrinsic absorption of column density $10^{22}\pcmsq$ (bottom, red). The lower panel shows the same as the middle one, but versus wavelength over the RGS band and with a constant 3 keV plasma spectrum (green). Identification and emissivity of major emission lines are given in \citep{Sanders2011}. }  
    \label{fig:my_label}
\end{figure}

The weakness of the soft X-ray emission expected from a cooling flow had been first noticed in proportional counter and solid-state spectrometer data 30 years ago and the lack of cool X-ray emitting gas was soon considered to be an effect of  X-ray absorption \citep{White91, Johnstone92, Fabian1994}. The effect was later found to be more subtle than just an absorbing screen around the cooling gas: an intrinsic absorber mixed with that gas is required \citep{Allen1997}. Detailed work on low resolution ASCA and ROSAT spectra showed that this can fit well and allow significant cooling flows \citep{Allen2000, Allen01}.  Here we call such  absorbed flows Hidden Cooling Flows (HCFs). 
For our first new paper on this topic \citep{Fabian22}, hereafter HCF1, we  fitted an intrinsic absorber model to RGS spectra of 3 cool core clusters -- Perseus, Centaurus and A1835 which are known to have plenty of cold gas but seem to lack evidence of cooling hot gas -- finding that significant absorbed cooling rates are statistically allowed at values of 1/2 to 1/4 of the unsuppressed values.

Example model spectra of unabsorbed, fully absorbed, truncated and intrinsically-absorbed cooling flows are shown in Fig. 1. The intrinsically-absorbed spectrum is intermediate between no and full absorption. The lower temperature components occur at lower energy and so are absorbed more. The intrinsic absorption model is the result of summing over a large number of interleaved emission and absorption layers. The absorption multiplier is given by $(1-\exp^{-\sigma N_{\rm H}})/(\sigma N_{\rm H})$, where $N_{\rm H}$ is the total absorbing column density and $\sigma$ is the energy-dependent absorption cross-section. 
We can envisage more complex models but have started with the above basic model which minimises free parameters. 

The absorbed luminosity is mostly from gas which cools below 1 keV, so for a 5 keV cluster it amounts to about 20 per cent of the total cooling flow luminosity. We find the luminosity to be  consistent with the Far InfraRed (FIR) luminosity of dusty gas in the galaxy core. So in principle there is not an energy problem. Mass is perhaps the main problem in the sense of  the implied accumulation of cold gas. The observed BCGs have considerable amounts of atomic gas (e.g. H$\alpha$ nebulae) and particularly cold molecular gas (e.g molecular emission from CO and other species), with total masses ranging from $10^{8}- 10^{11}\Msun$.  They also have large visible stellar masses of about $10^{12}\Msun$ or more. The expected accumulation of cooled gas over one billion years from a hidden cooling flow can  however be one to two orders of magnitude larger than the already-observed cold diffuse gas mass. This is known as the "cooling flow problem", see e.g. \citep{McDonald2018}.  HCF1 discusses several ways of overcoming this problem including a) the observed cold gas being only a small fraction of the total with much of it being so cold as to be undetectable; b) AGN feedback dragging cold gas clouds to larger radii so spatially redistributing the cooled gas and explaining the abundance profiles; and/or c) cloud fragmentation and collapse into low mass stars, brown dwarfs and planetoids in the high pressure environment of cool core. Note that the solutions are not mutually exclusive.

We note too that there is not a space problem since if gas within the innermost 50 kpc of a 5 keV cluster cools to 5~K, then its density changes by $10^7$ and it would all fit within 100~pc (at the current pressure). If it forms gravitationally-bound low mass objects, its volume is then of course enormously less.  

Here we study a wider sample of cool cores starting with a group, then 7 more clusters and 2 elliptical galaxies, complementing the 3 clusters already studied (Perseus, Centaurus and A1835). Most have lower Galactic column densities $N_{\rm H}<10^{21}\pcmsq$, two are at redshift $z\sim 0.35$ and one is at $z=0.59$. The last three all have extreme properties. All of them statistically allow significant Hidden Cooling Flows.

\begin{table*}
  \caption{Observed targets, giving the used source position (deg; J2000), observation identifiers, extraction region (in per cent of the cross-dispersion PSF), background type (extraction region in per cent or model created using \textsc{rgsbkgmodel}) and average cleaned exposure of the RGS cameras.}
  
  \begin{tabular}{lrr>{\raggedright}p{5cm}rrr}
  Target & RA & Dec & OBSIDs & Region & Background & Exposure (ks) \\ \hline
  NGC5044 & $198.8498$ & $-16.3853$ & 0037950101 0554680101 & $95$ & Model & $138$ \\
  Sersic 159 & $348.4947$ & $-42.7254$ & 0123900101 0147800101 & $95$ & Model & $141$ \\
  A262 & $28.1930$ & $36.1526$ & 0109980101  0109980601  0504780101  0504780201 & $95$ & Model & $191$ \\
  A2052 & $229.1854$ & $7.0215$ & 0109920101   0109920301  0401520301  0401520501  0401520601  0401520801  0401520901  0401521101  0401521201  0401521301  0401521601  0401521701 & $95$ & Model & $146$ \\
  RXJ0821 & $125.2586$ & $7.8636$ & 0880810101  0880810201  0880810301  0880810401  0880810501  0880810601 & $95$ & $99$ & $239$ \\
  RXJ1532 & $233.2242$ & $30.3494$ & 0039340101 0651240101 & $95$ & $99$ & $45$ \\
  MACS1931 & $292.9567$ & $-26.5760$ & 0693180101 & $95$ & $99$ & $40$ \\
  Phoenix cluster & $356.1832$ & $-42.7202$ & 0693661801 0722700101 0722700201 & $95$ & $99$ & $242$ \\
  M84 & $186.2656$ & $12.8870$ & 0673310101 0821800101 0821800201 0821800301 & $95$ & Model & $456$ \\
  M49 & $187.4448$ & $8.0005$ & 0112550601 0200130101 & $95$ & Model & $101$ \\
  \hline
  \end{tabular}
\end{table*}

\begin{figure*}
    \centering    \includegraphics[width=0.36\textwidth]{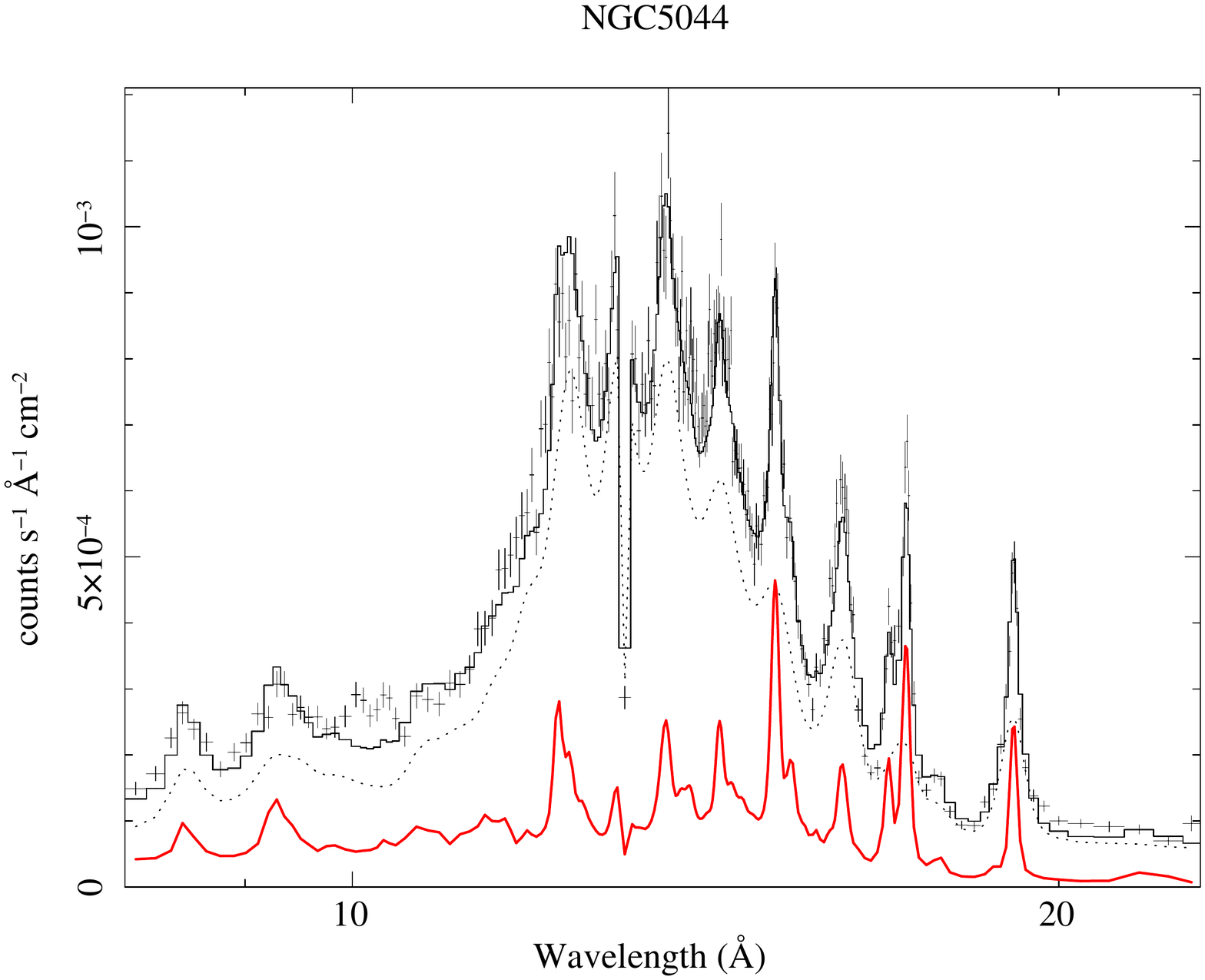}
    \hspace{-0.85cm}\includegraphics[width=0.36\textwidth]
    {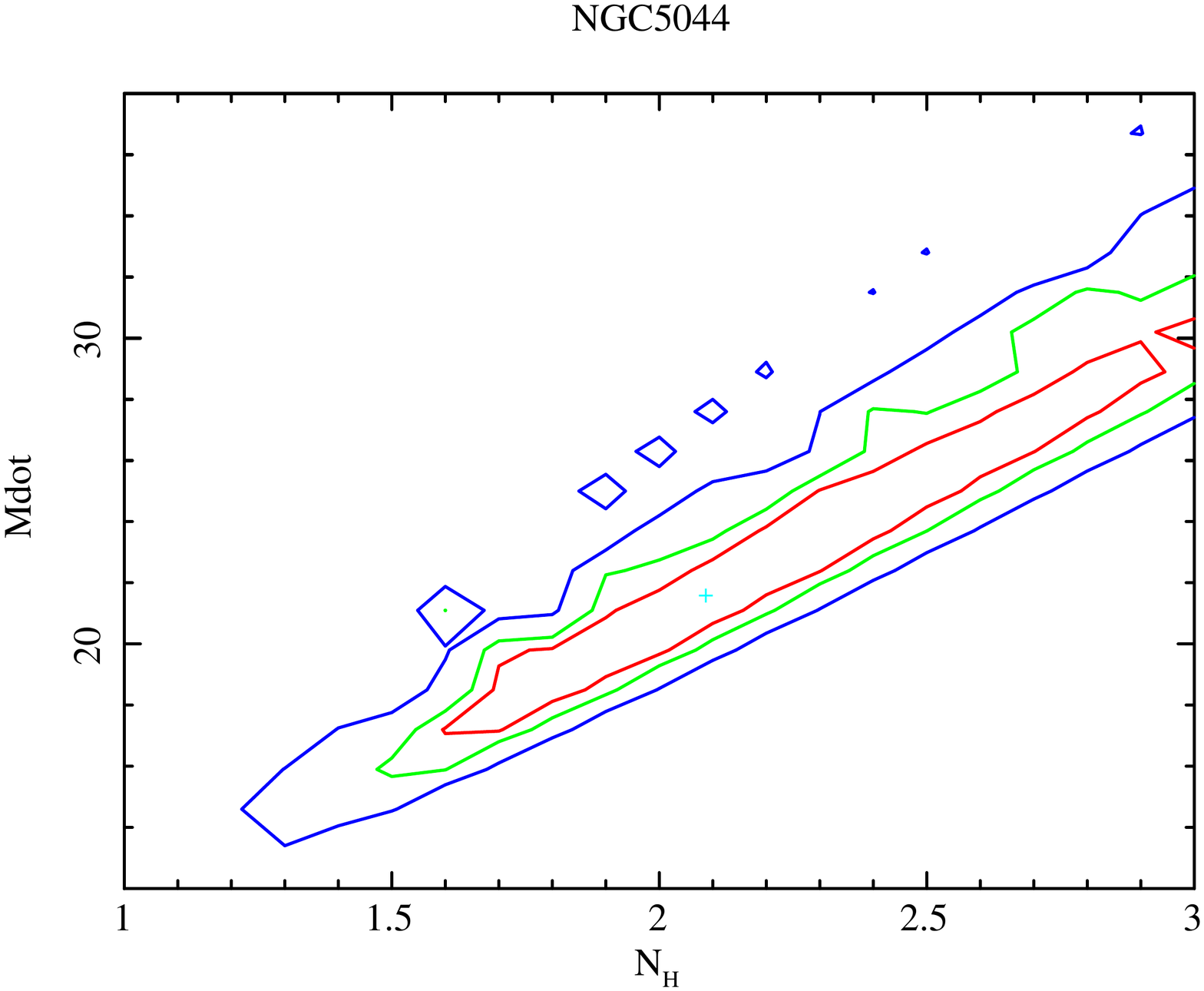}
      \hspace{-0.85cm} \includegraphics[width=0.36
      \textwidth]{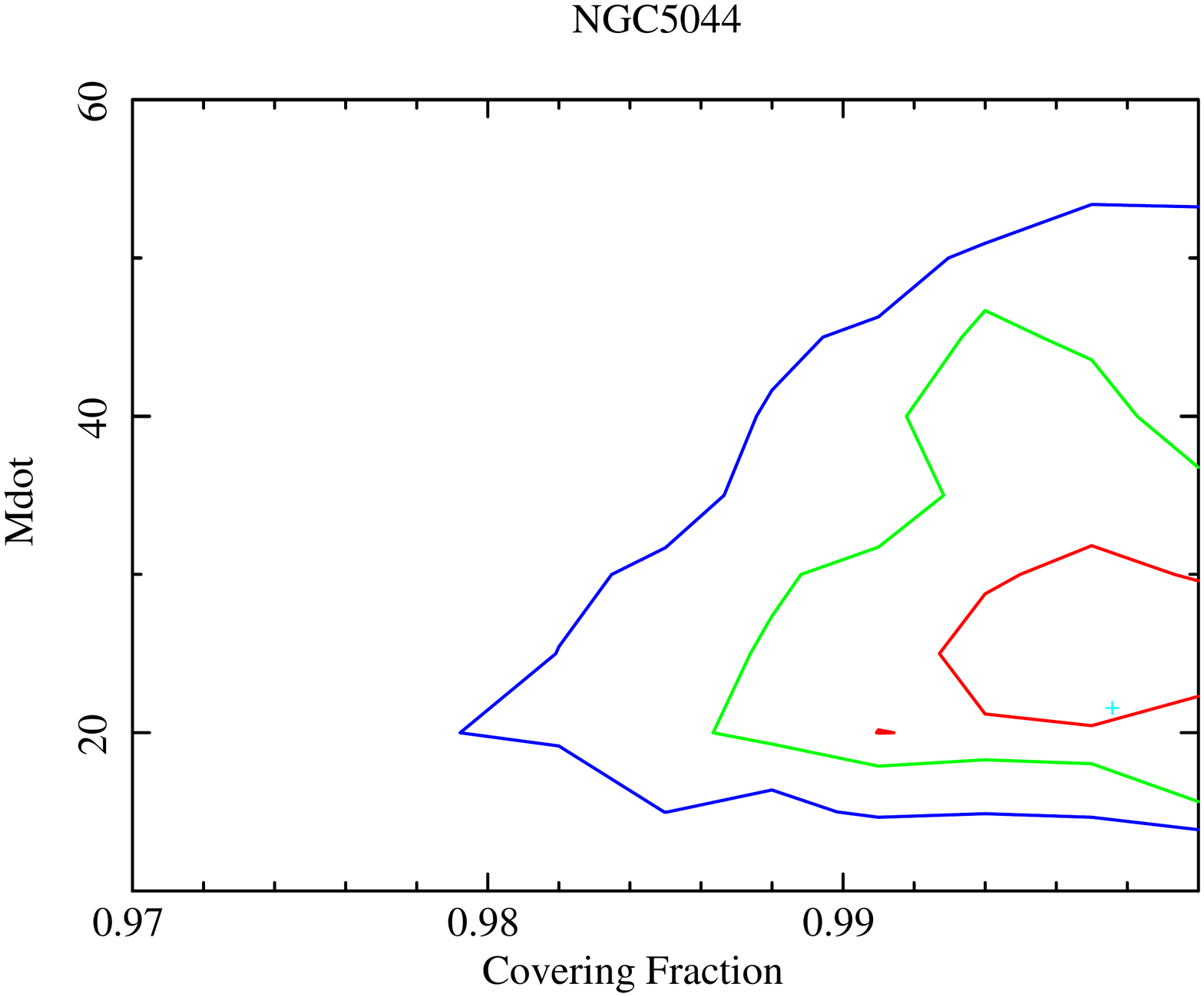}
    \caption{ Left to Right: RGS spectrum of NGC5044 with HCF component shown in red, Mass cooling rate in $\Msunpyr$ versus  total column density in units of $10^{22}\cmsq$, Mass cooling rate versus Covering Fraction of the HCF component. Contours at 68\% (red), 90\% (green) and 99\% (blue). }
\end{figure*}

\begin{figure*}
    \centering
    \includegraphics[width=0.36\textwidth]{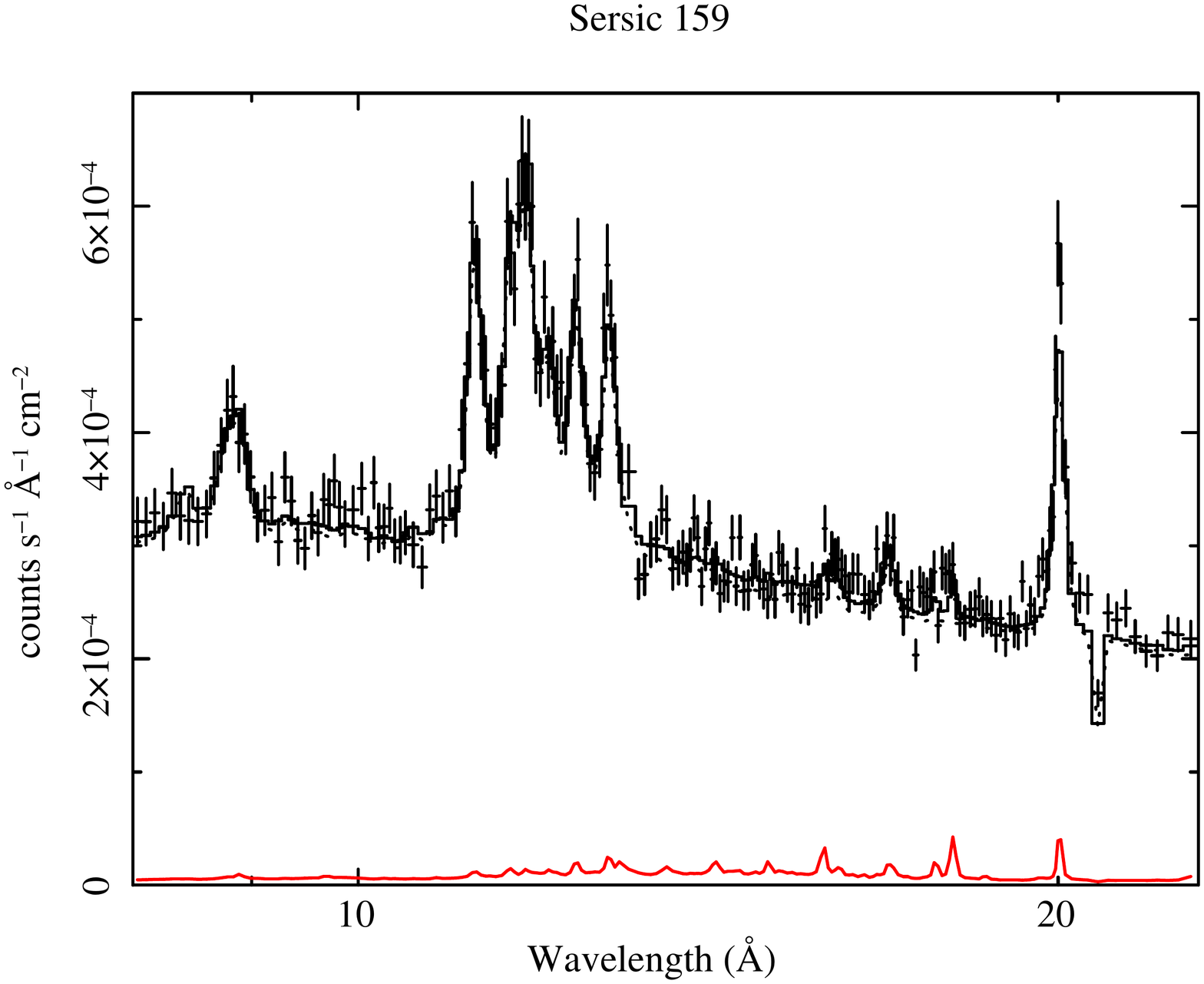}
       \hspace{-0.85cm}  \includegraphics[width=0.36\textwidth]{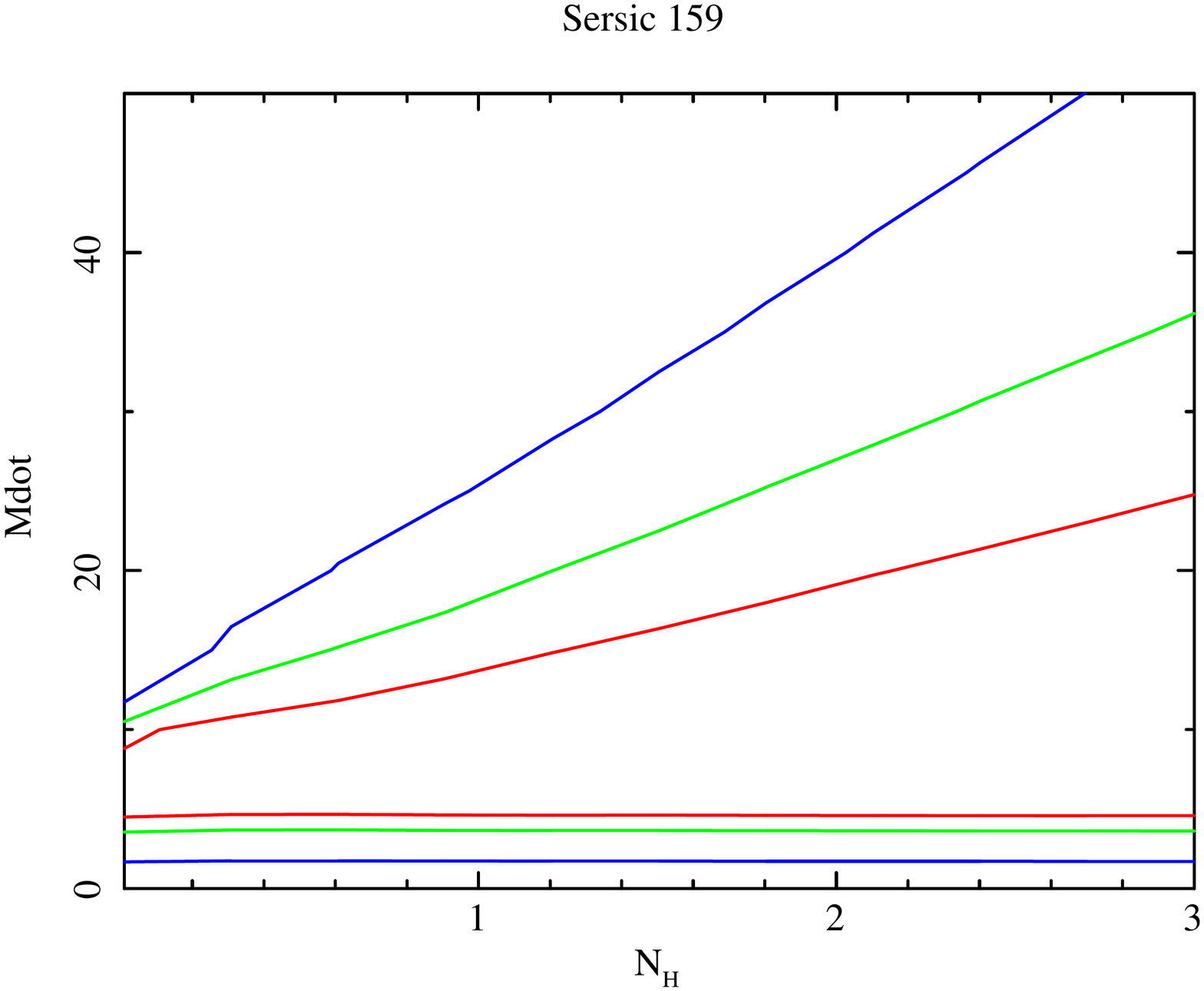}
    \hspace{-0.85cm}  \includegraphics[width=0.36\textwidth]{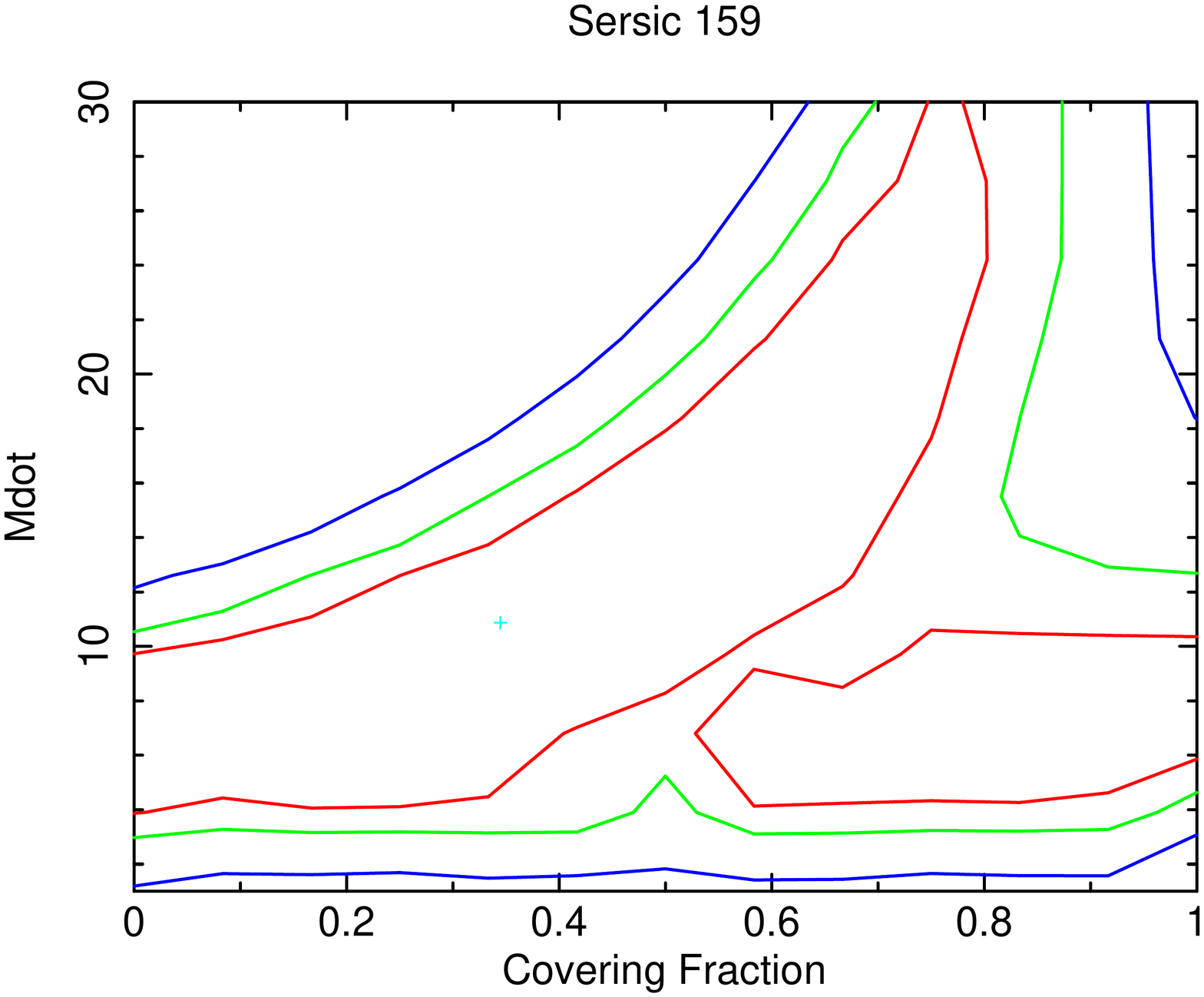}
    \caption{Sersic 159, with details as in Fig. 2.}
    \label{fig:}
\end{figure*}

\begin{figure*}
    \centering
    \includegraphics[width=0.36\textwidth]{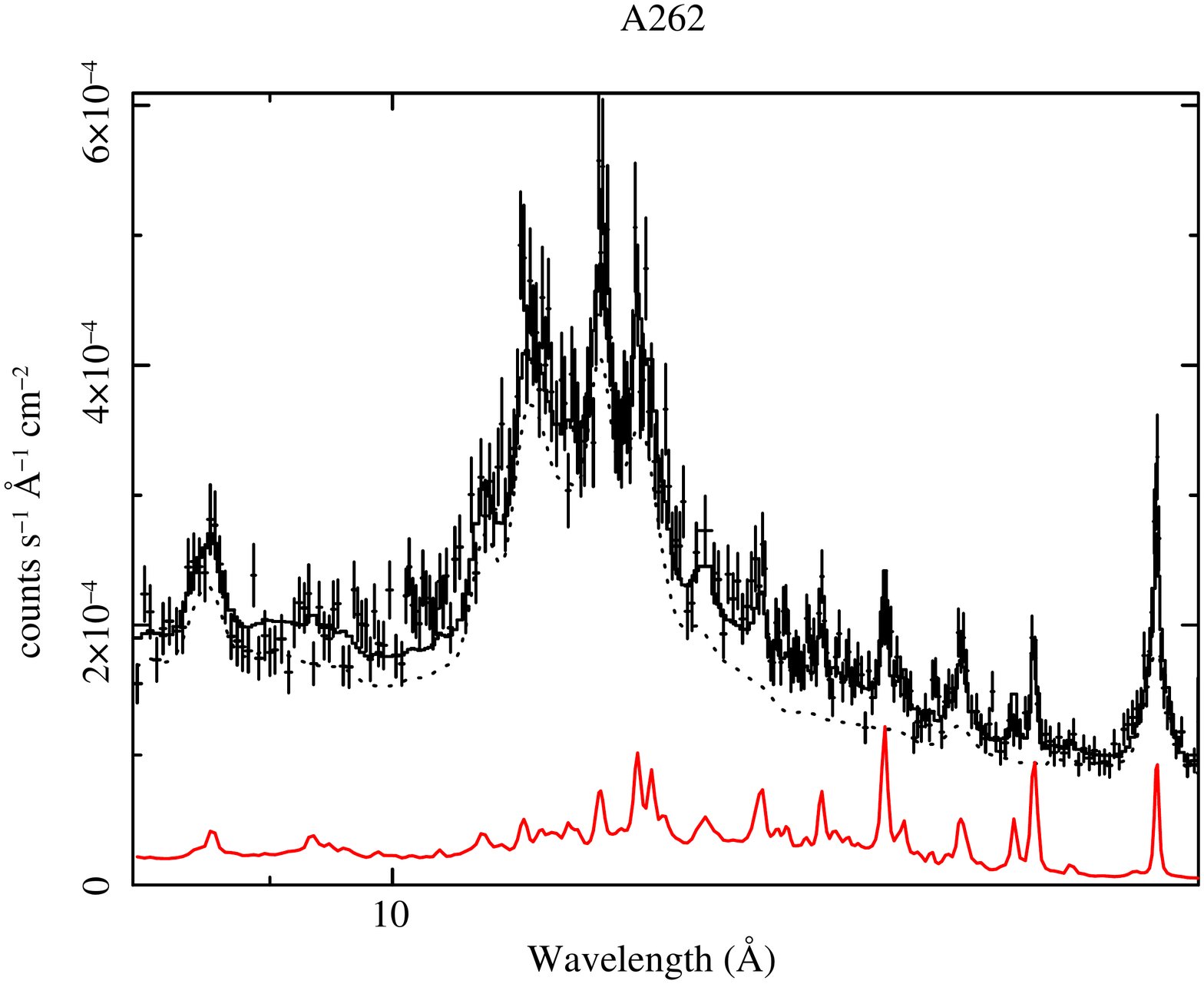}
      \hspace{-0.85cm}   \includegraphics[width=0.36\textwidth]{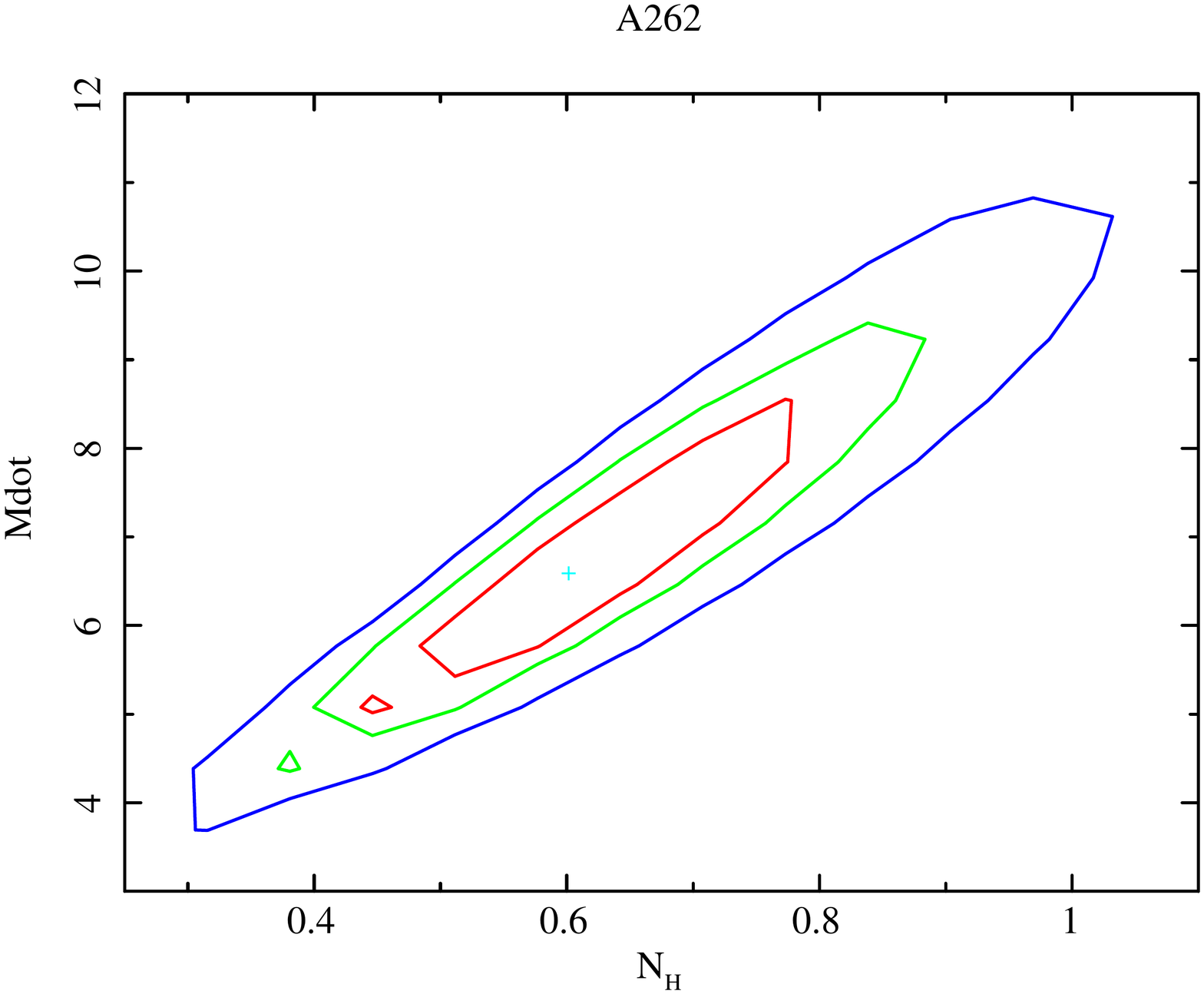}\hspace{-0.85cm} 
     \includegraphics[width=0.36\textwidth]{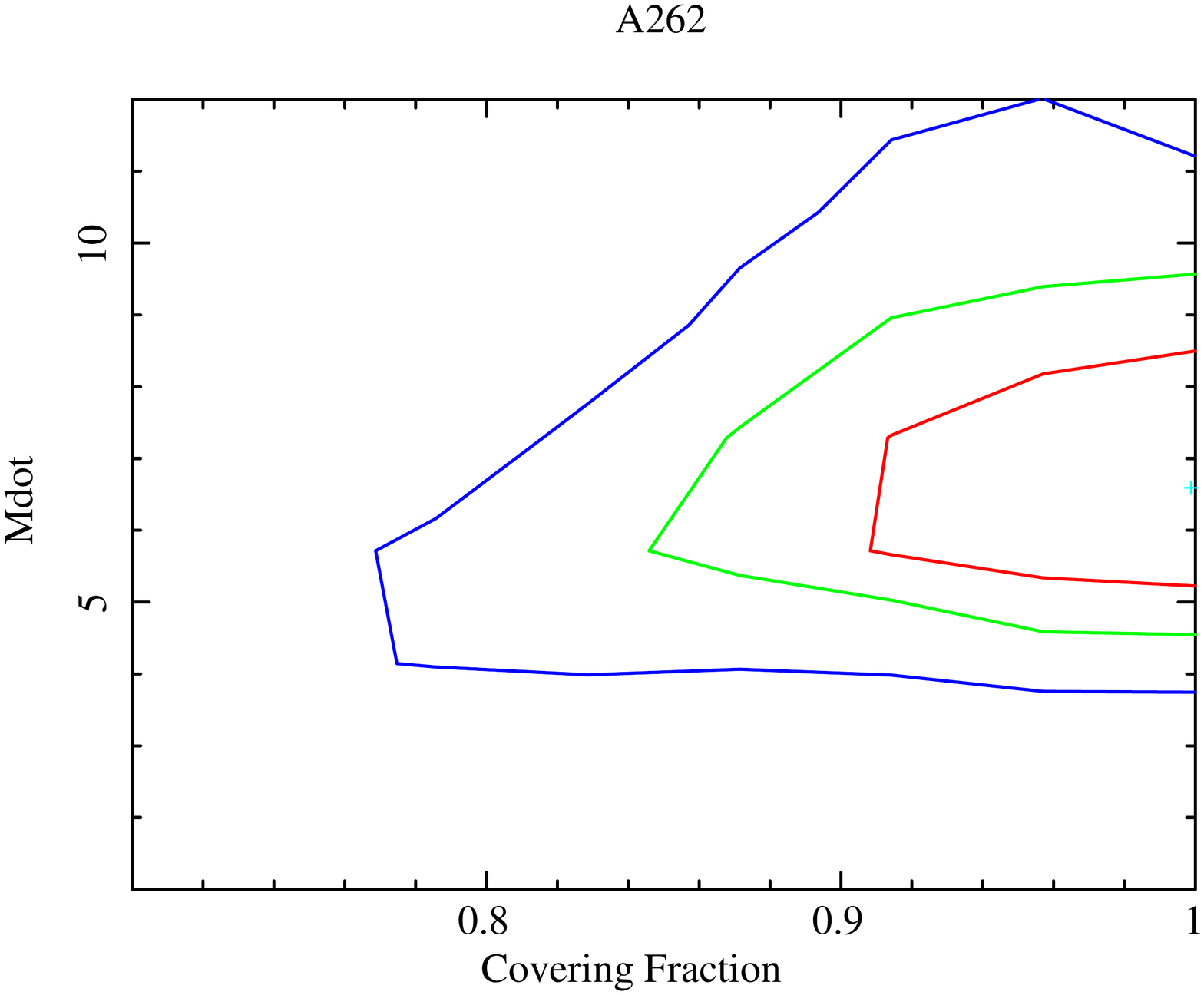}
    \caption{A262, with details as in Fig. 2.
   }
    \label{fig:my_label}
\end{figure*}

\begin{figure*}
    \centering
    \includegraphics[width=0.36\textwidth]{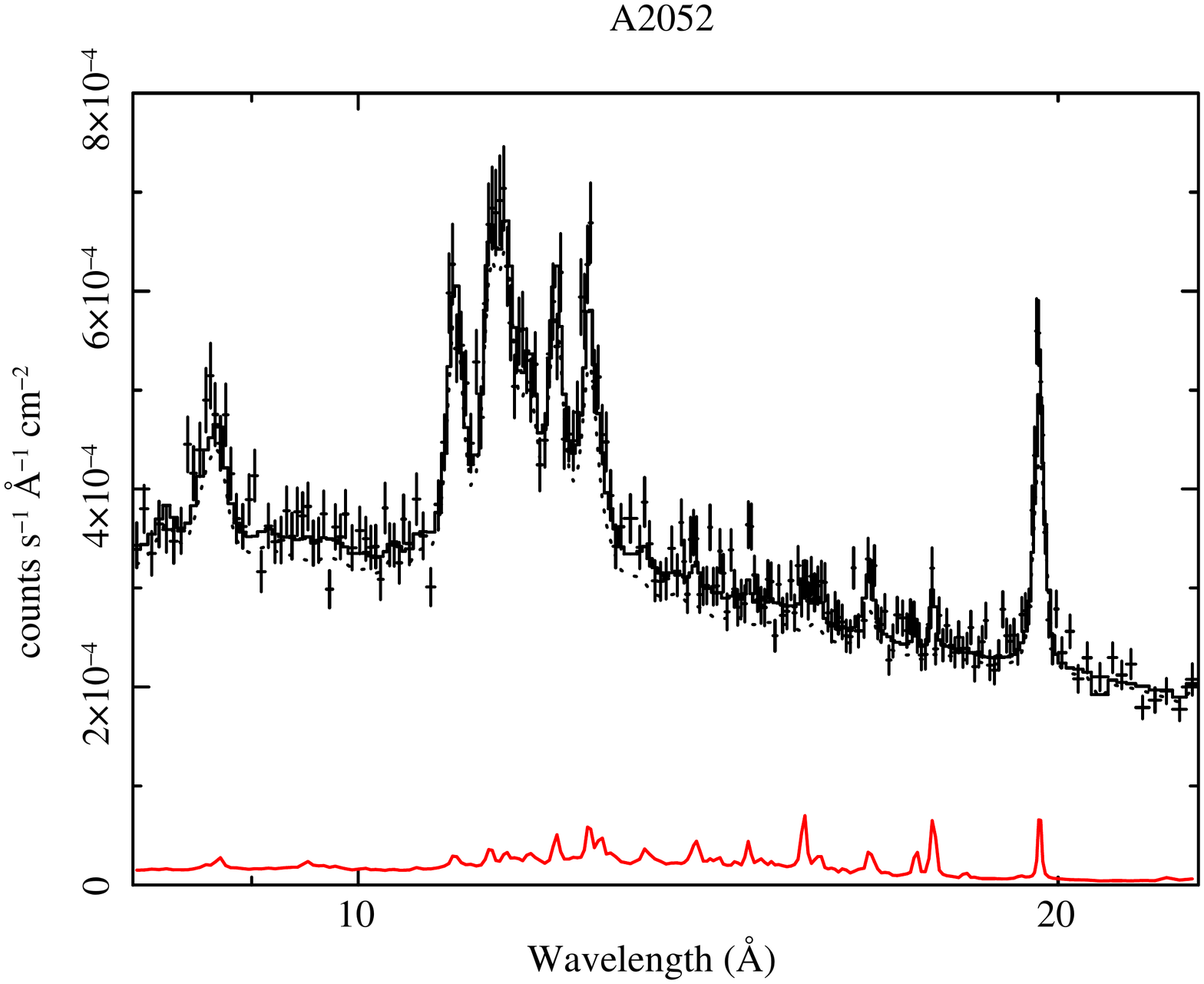}
      \hspace{-0.85cm}   \includegraphics[width=0.36\textwidth]{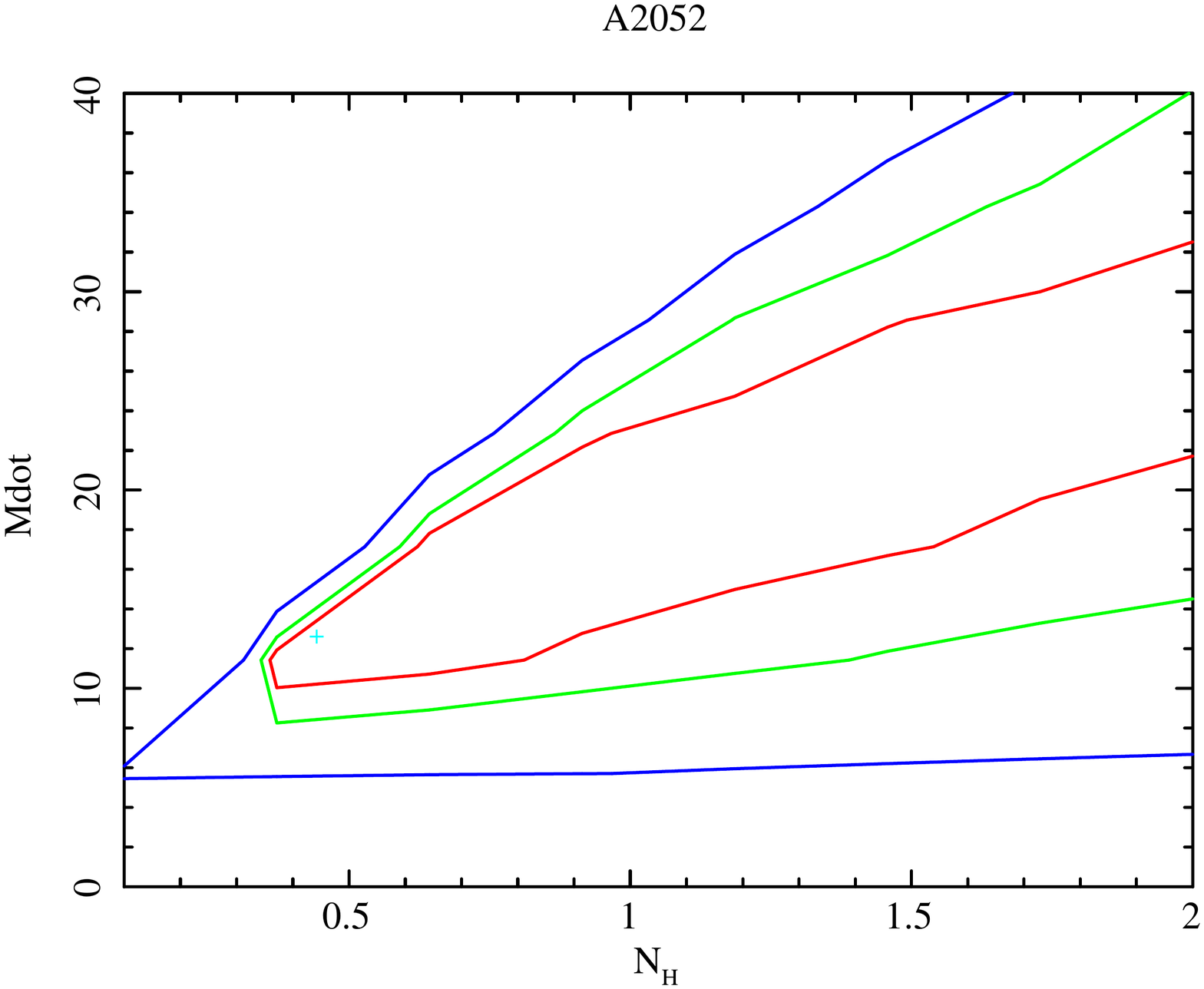}\hspace{-0.85cm} 
    \includegraphics[width=0.36\textwidth]{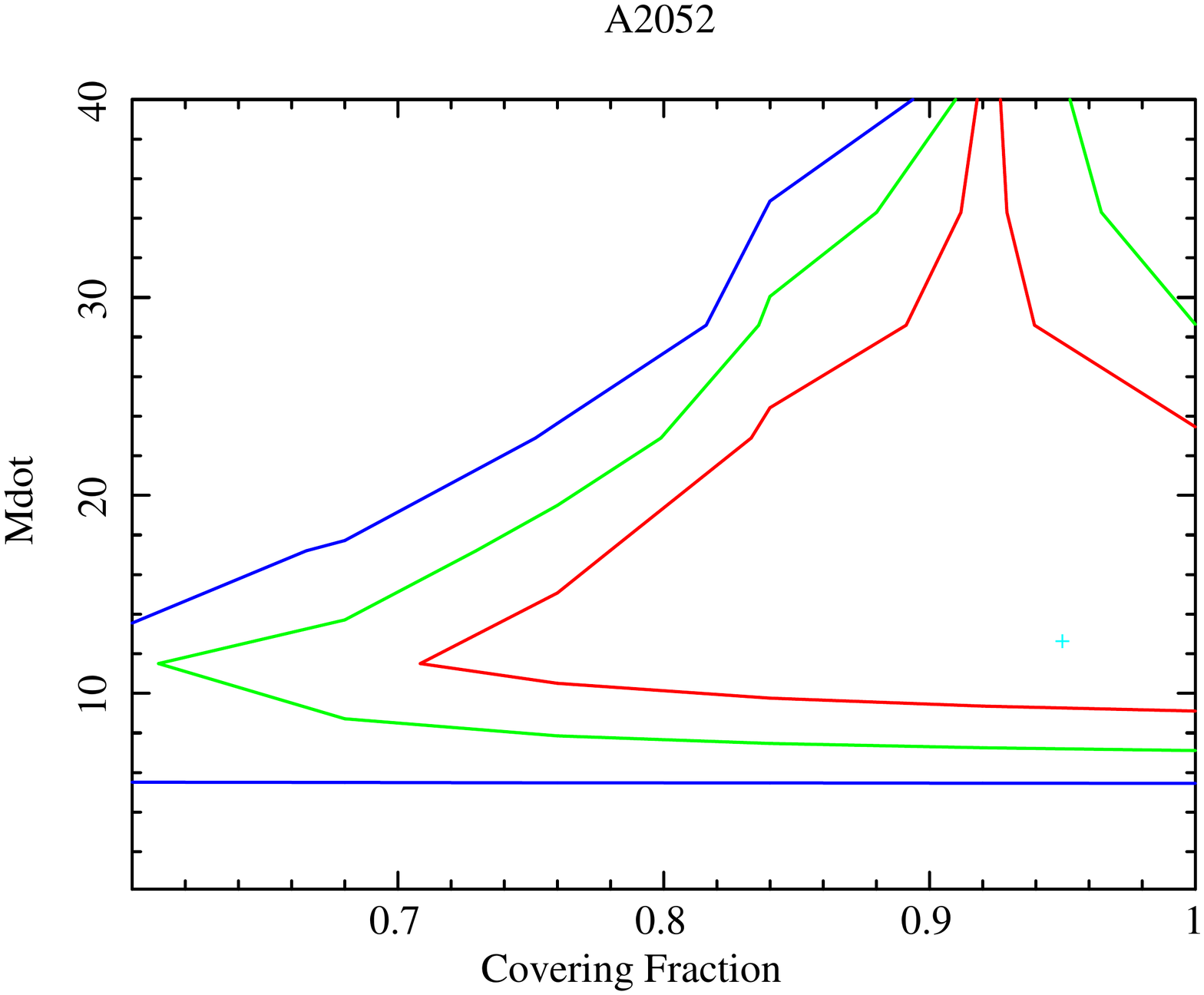}
    \caption{A2052, with details as in Fig. 2. }
\end{figure*}

\begin{figure*}
    \centering
    \includegraphics[width=0.36\textwidth]{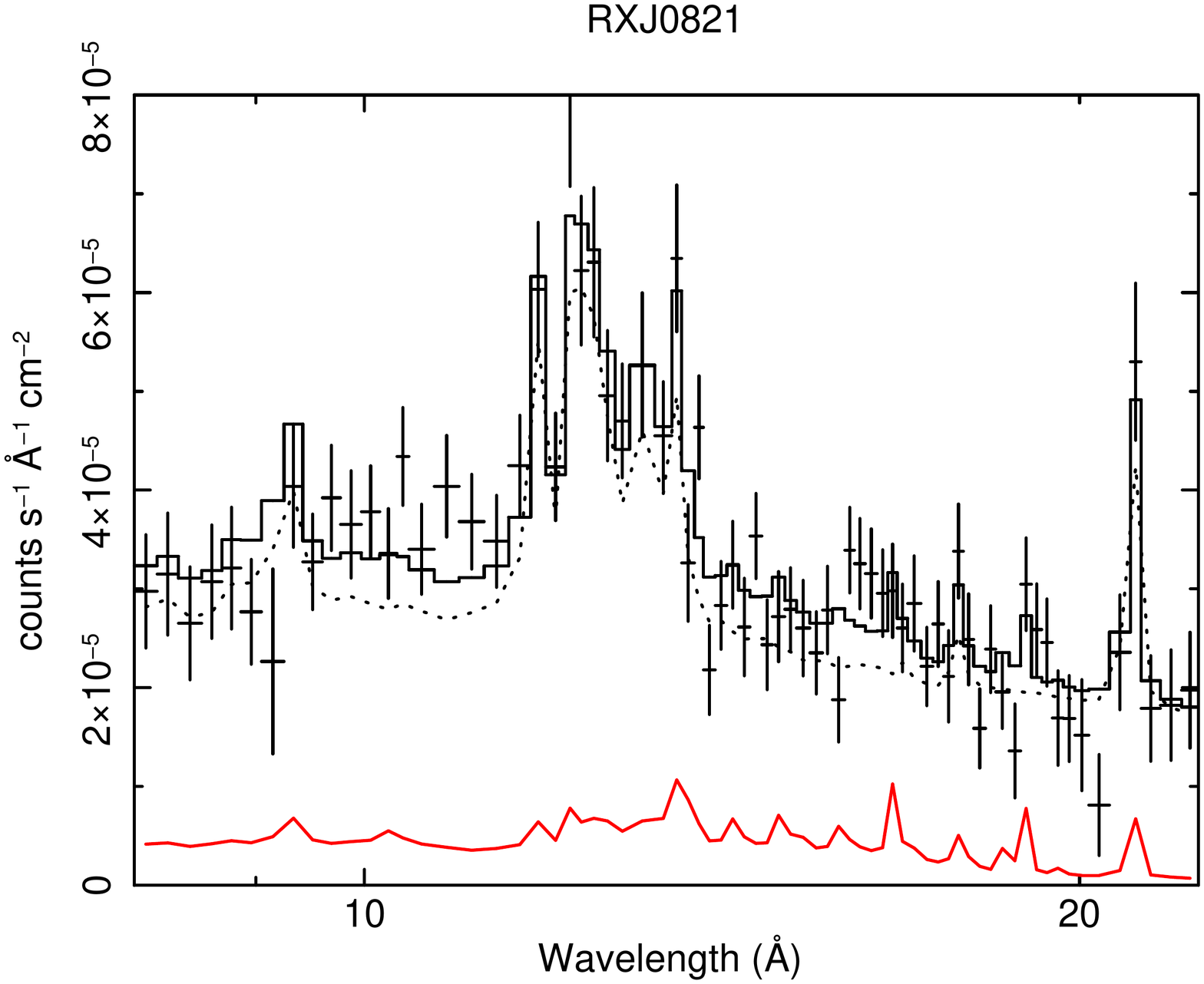}
       \hspace{-0.85cm}  \includegraphics[width=0.36\textwidth]{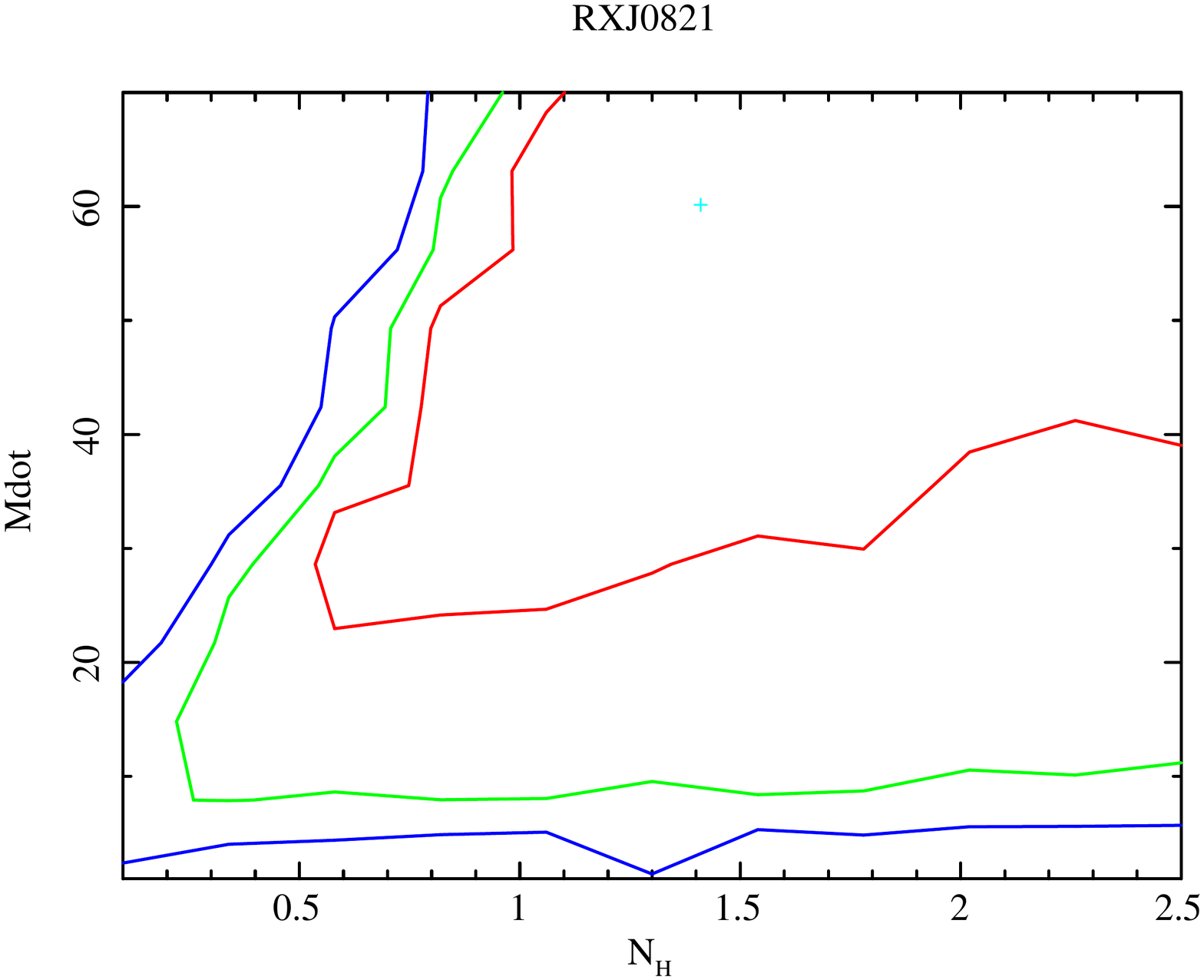}\hspace{-0.85cm} 
    \includegraphics[width=0.36\textwidth]{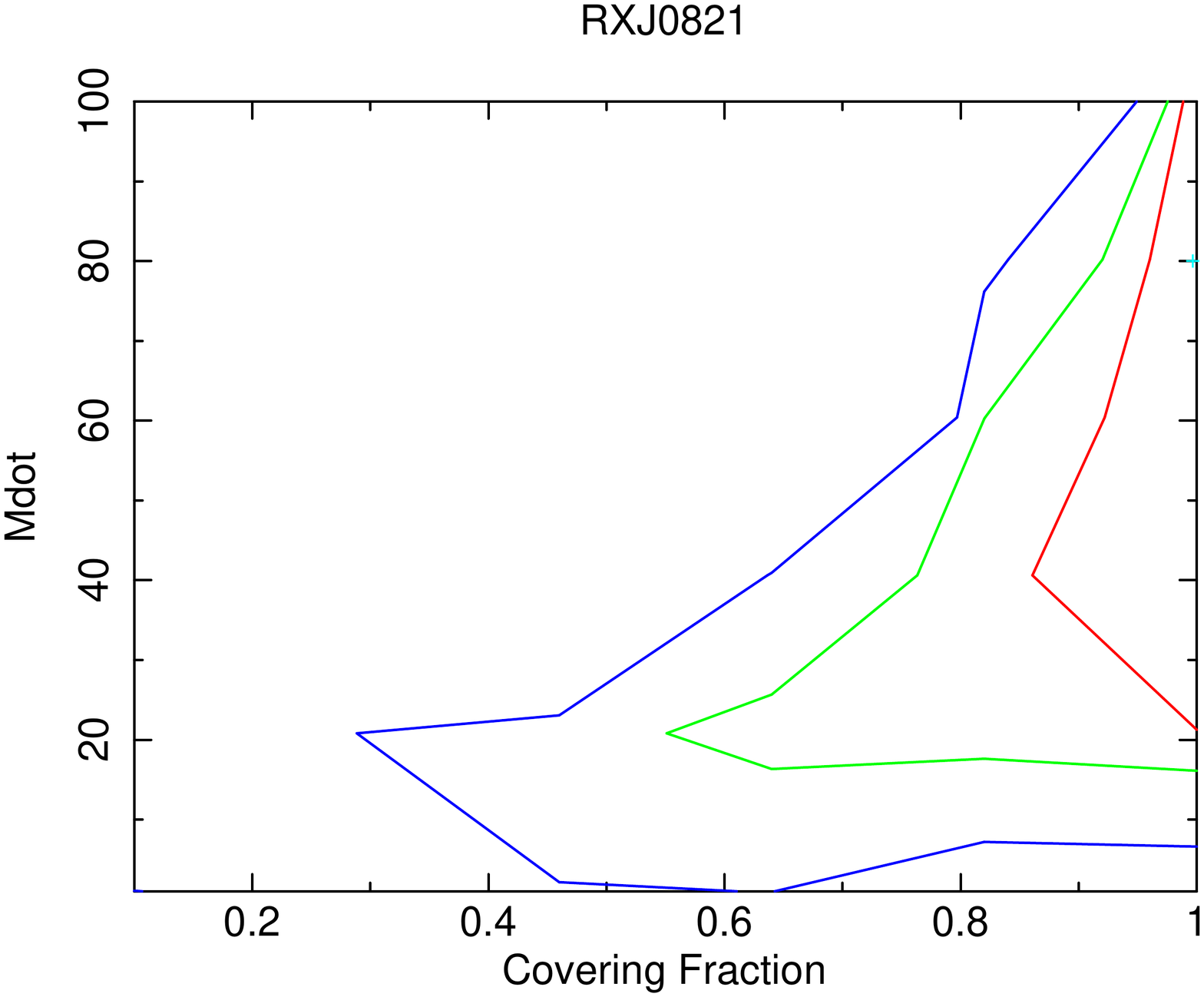}
    \caption{RXJ0821, with details as in Fig. 2. }
\end{figure*}

\begin{figure*}
    \centering    \includegraphics[width=0.36\textwidth]{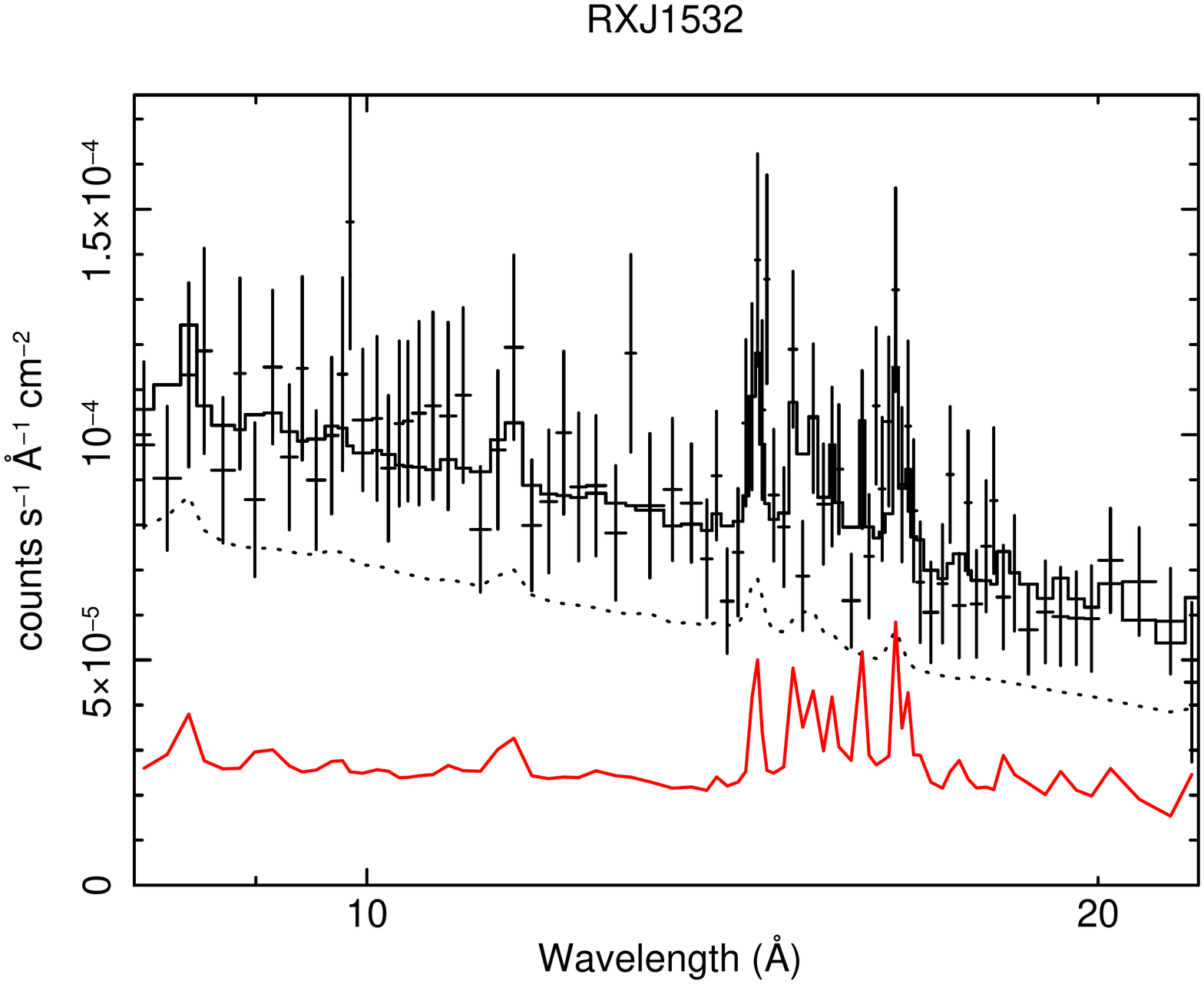}
    \hspace{-0.85cm}\includegraphics[width=0.36\textwidth]{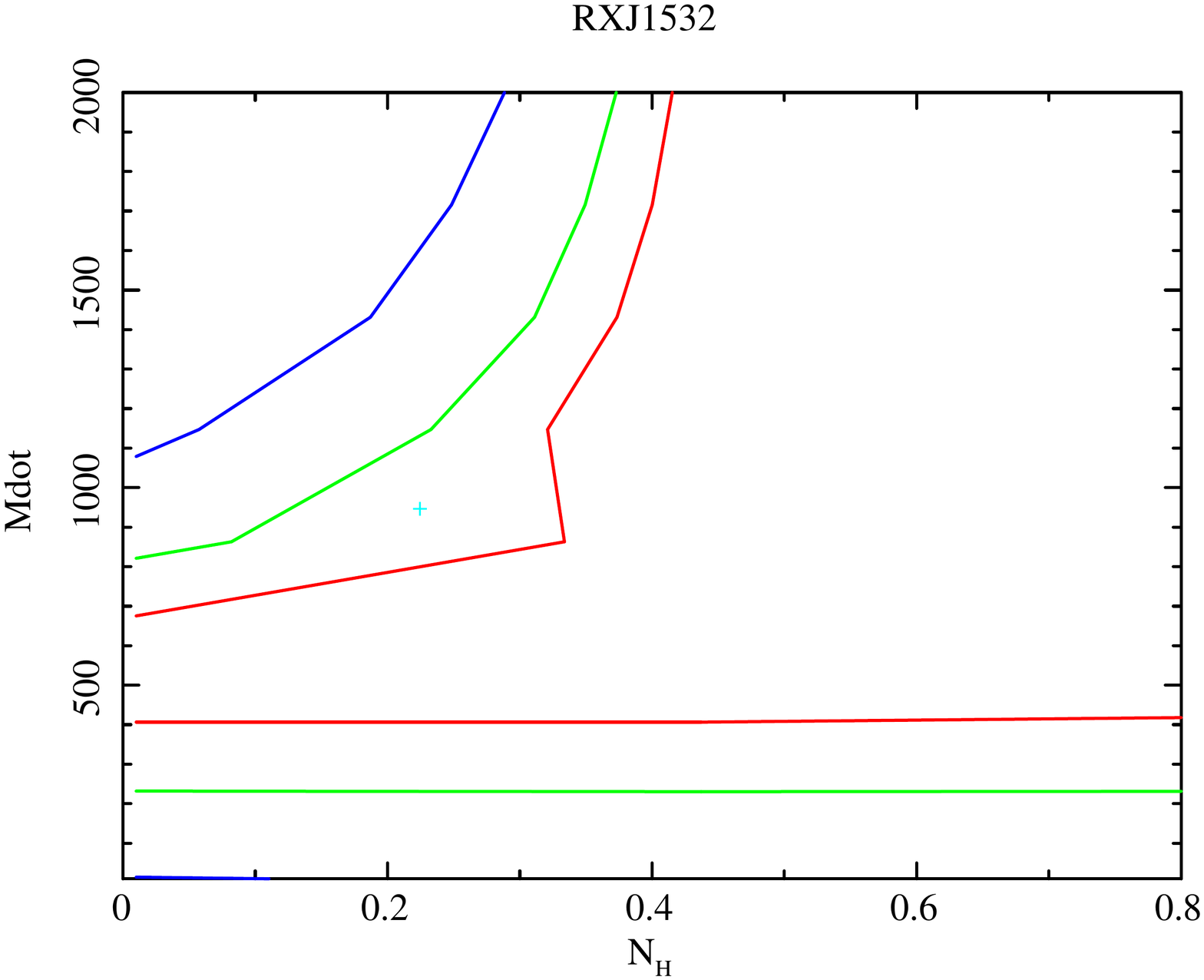}
      \hspace{-0.85cm} \includegraphics[width=0.36
      \textwidth]{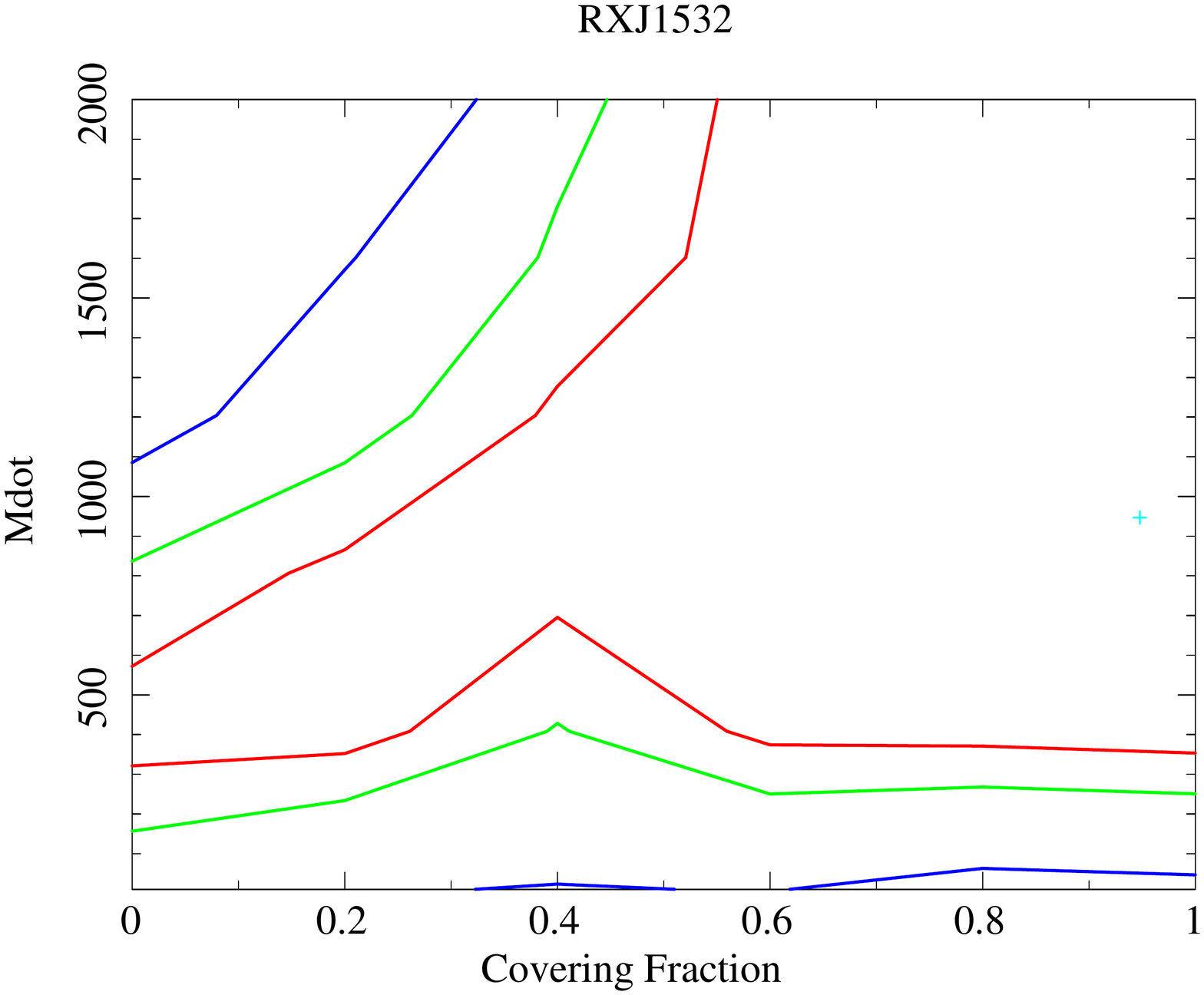}
    \caption{RXJ1532, with details as in Fig. 2. }
\end{figure*}

\begin{figure*}
    \centering
    \includegraphics[width=0.36\textwidth]{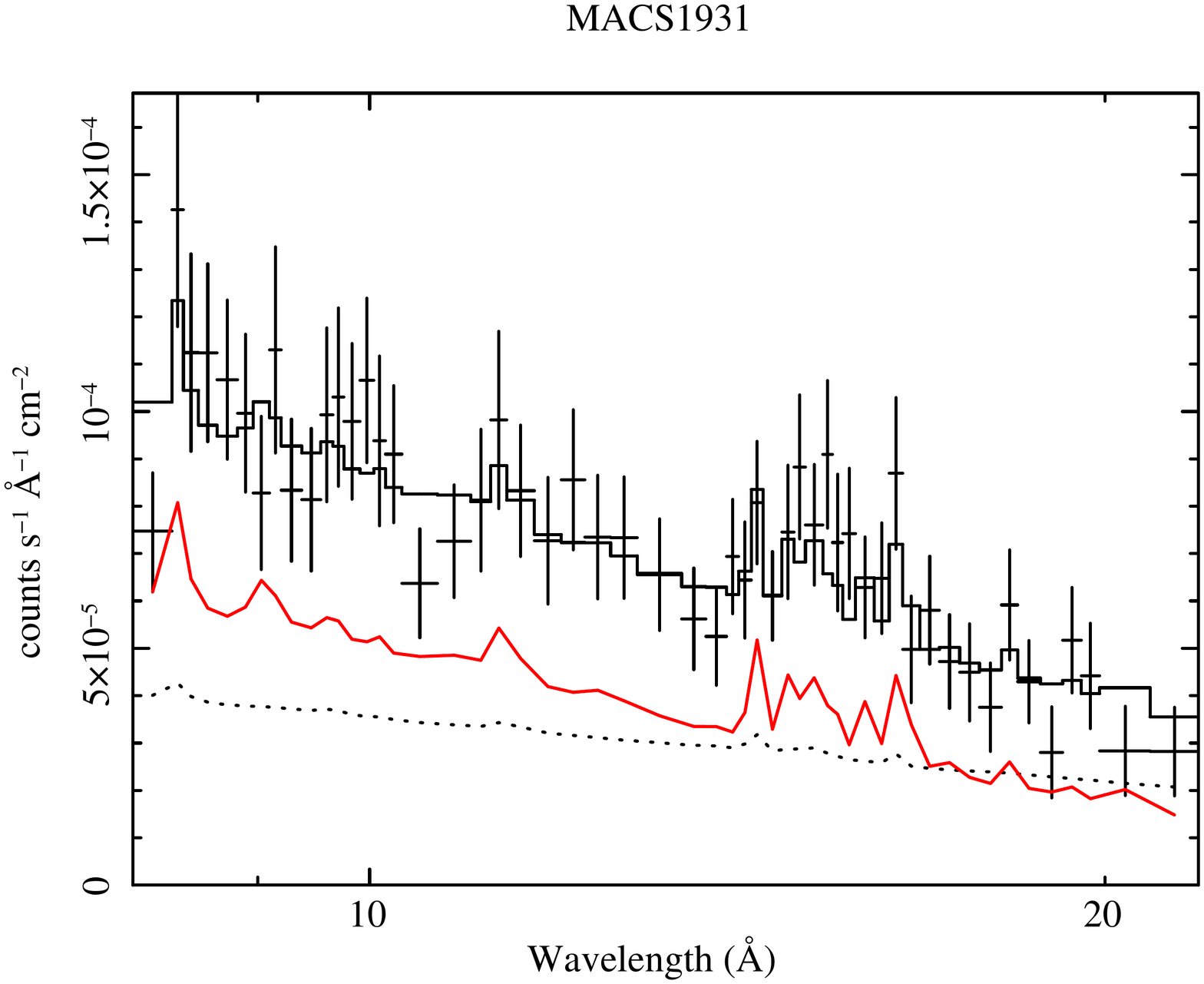}
        \hspace{-0.85cm}\includegraphics[width=0.36\textwidth]{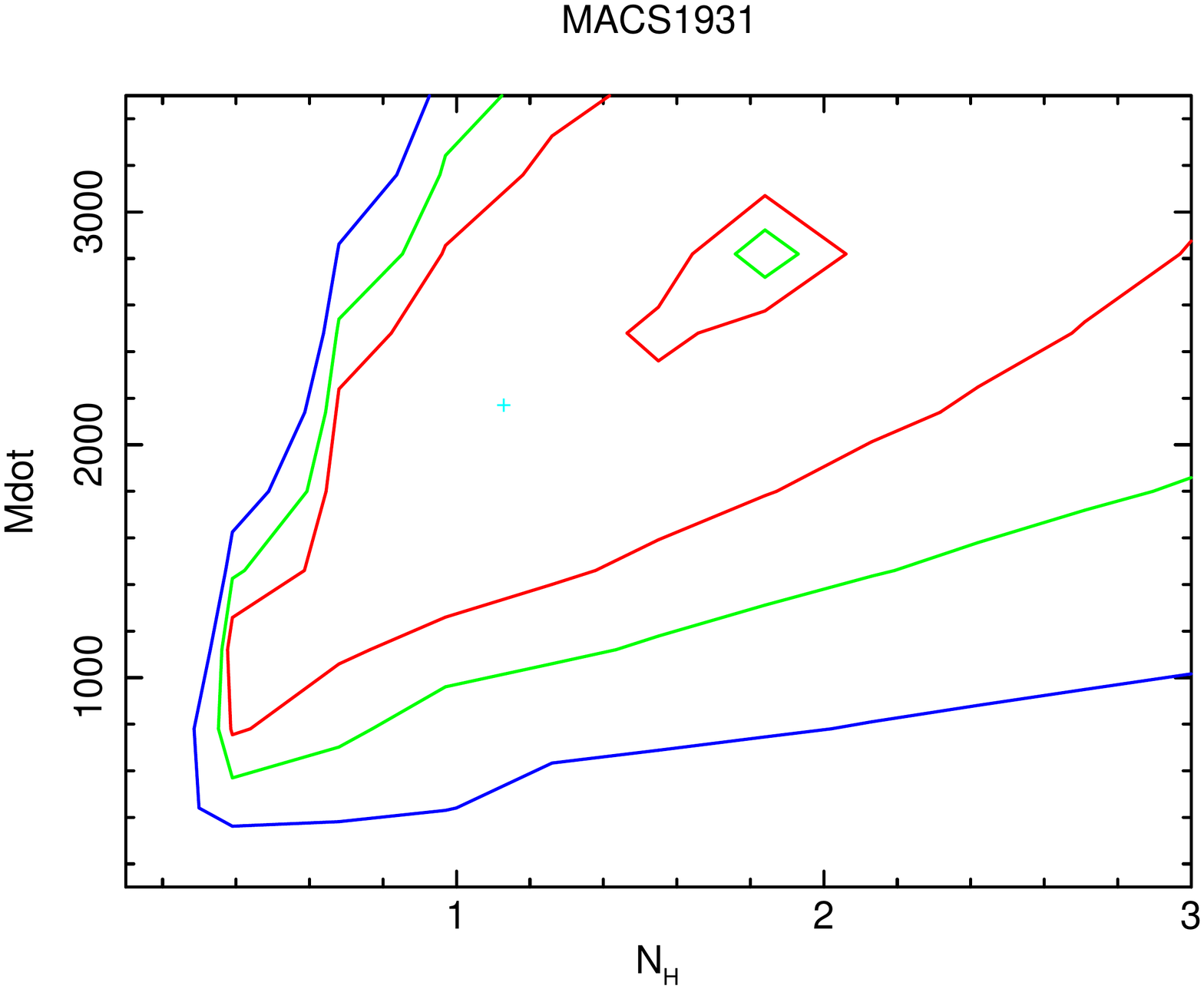}\hspace{-0.85cm}
    \includegraphics[width=0.36\textwidth]{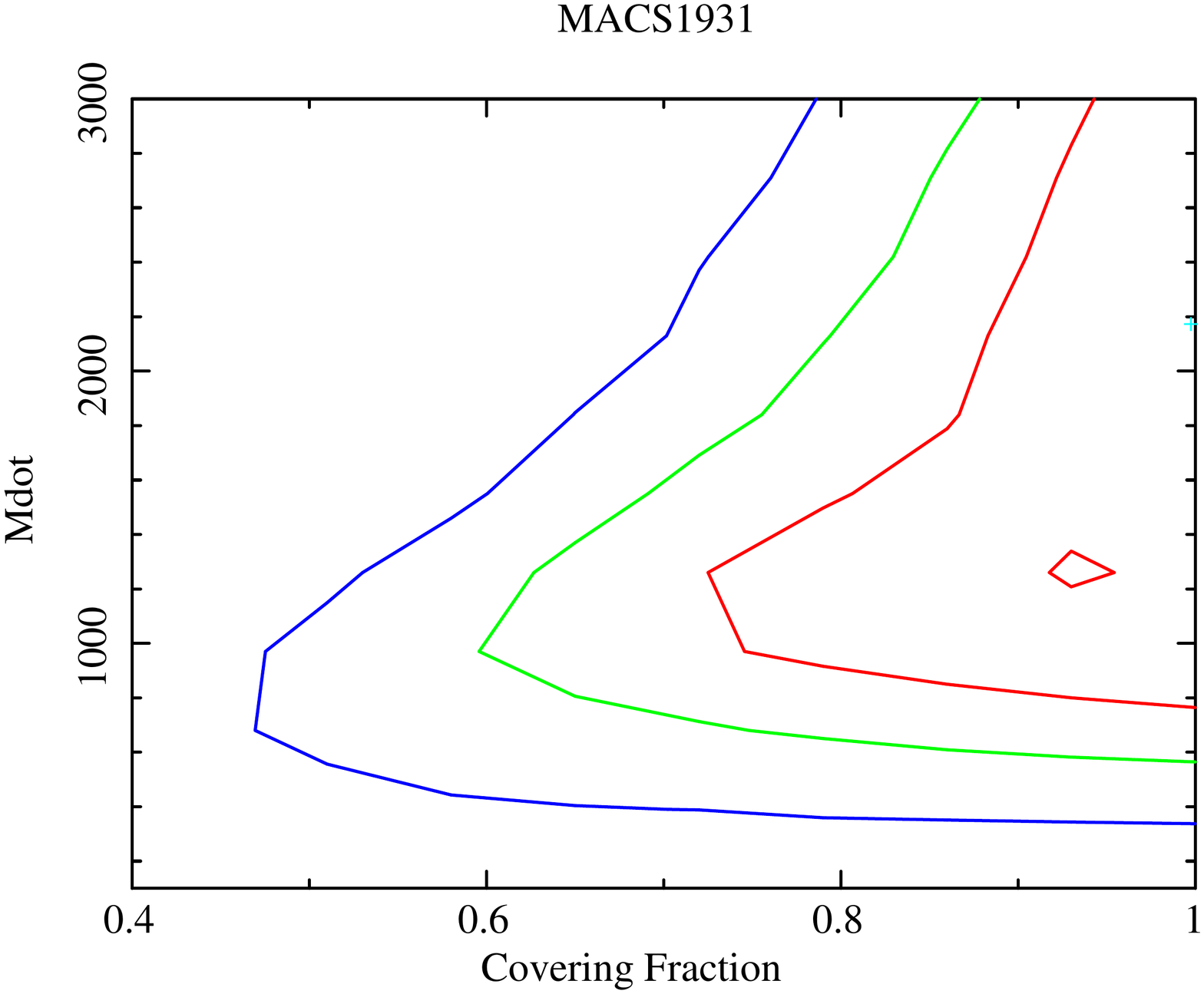}
    \caption{MACS1931, with details as in Fig. 2. }
\end{figure*}

\begin{figure*}
    \centering
    \includegraphics[width=0.36\textwidth]{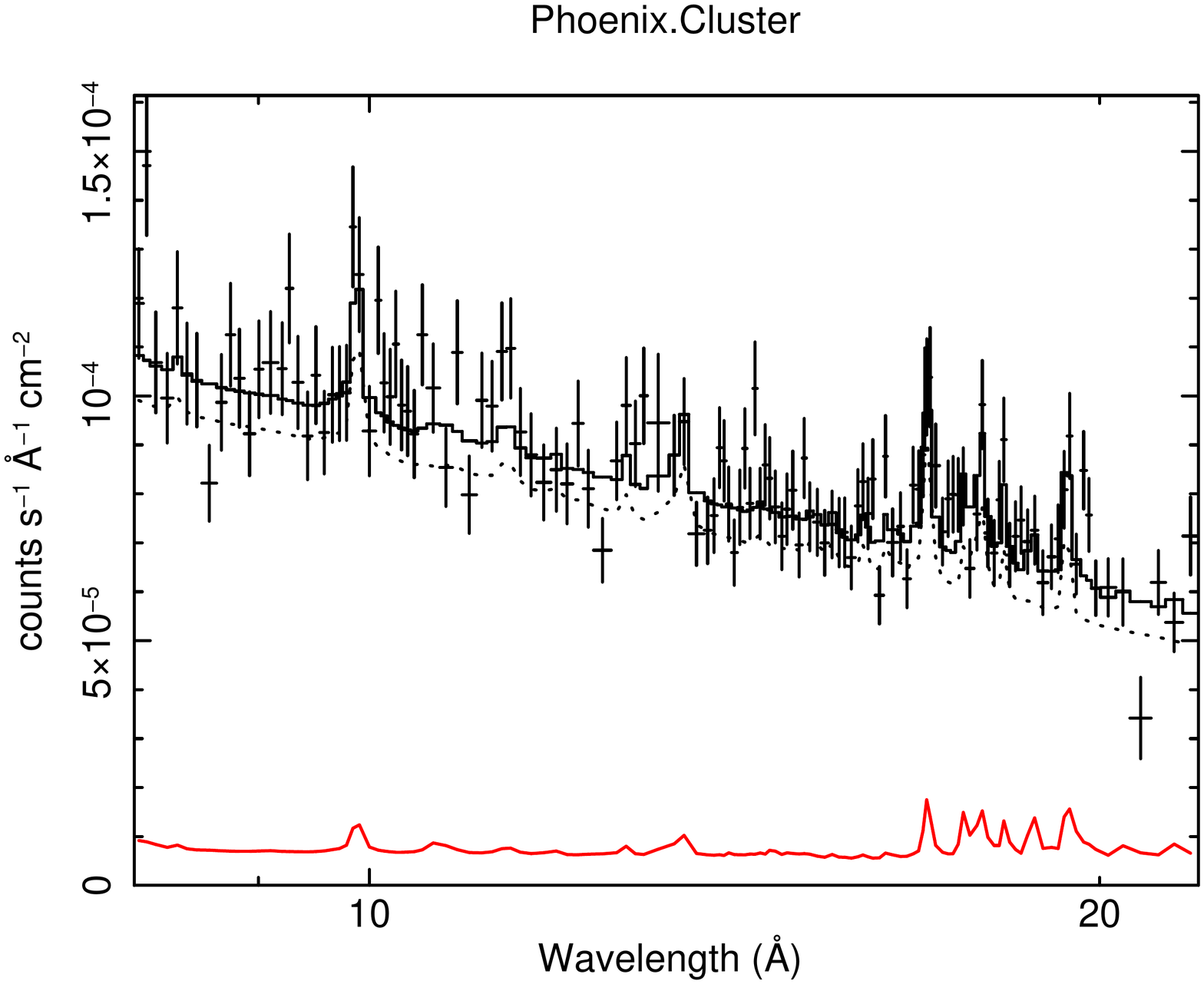}
        \hspace{-0.85cm}\includegraphics[width=0.36\textwidth]{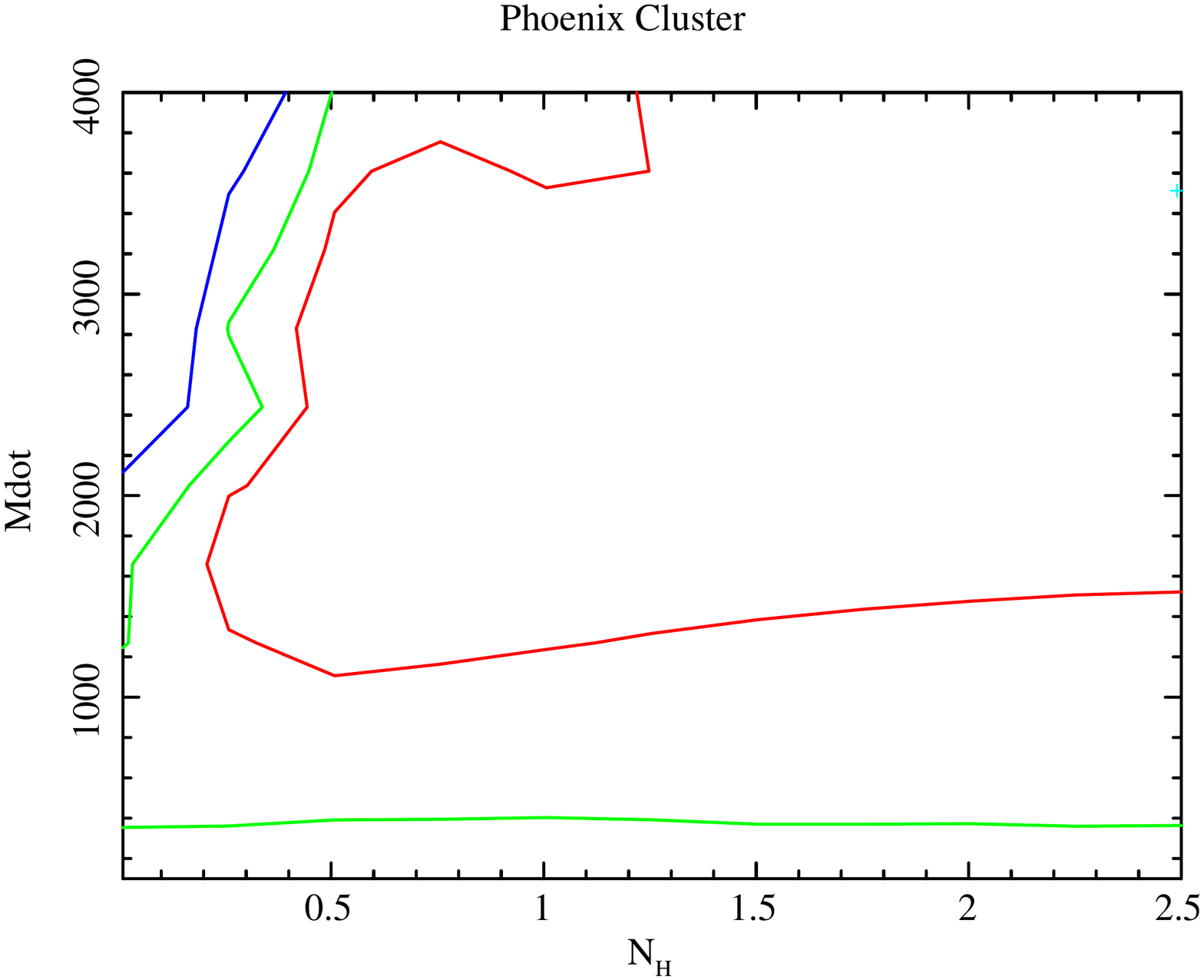}\hspace{-0.85cm}
    \includegraphics[width=0.36\textwidth]{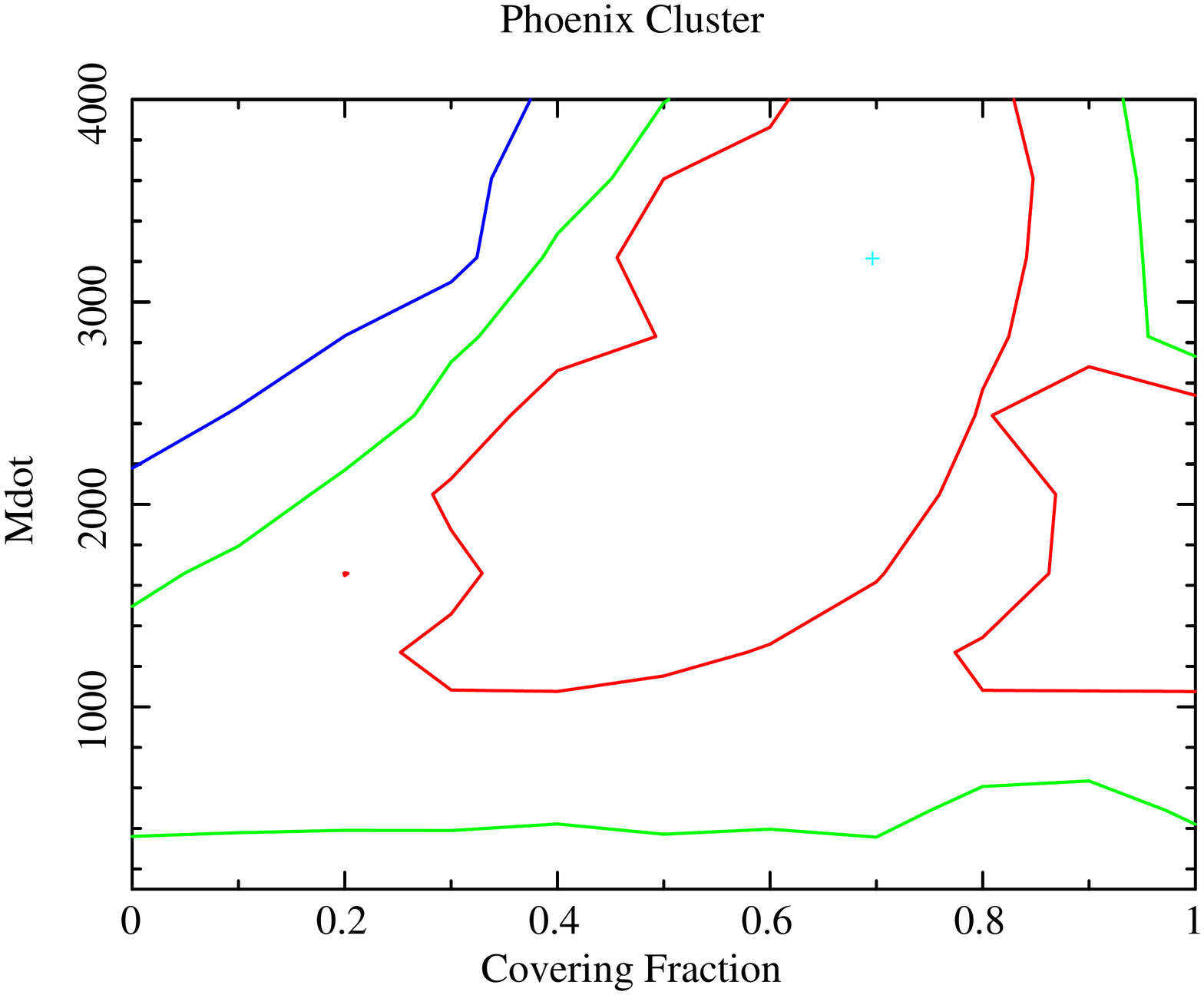}
    \caption{Phoenix Cluster, with details as in Fig. 2. }
\end{figure*}

\begin{figure*}
    \centering
    \includegraphics[width=0.36\textwidth]{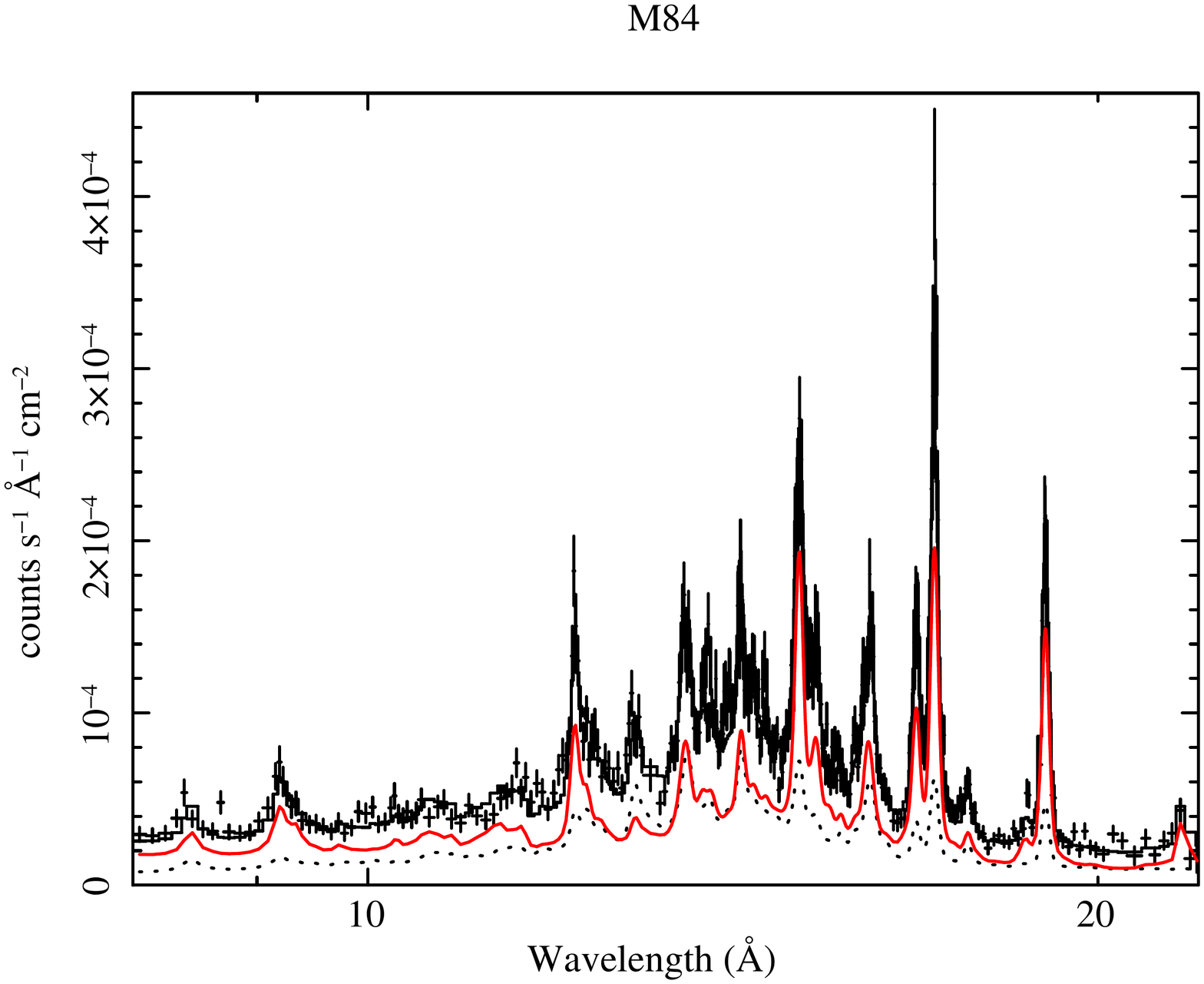}
        \hspace{-0.85cm}\includegraphics[width=0.36\textwidth]{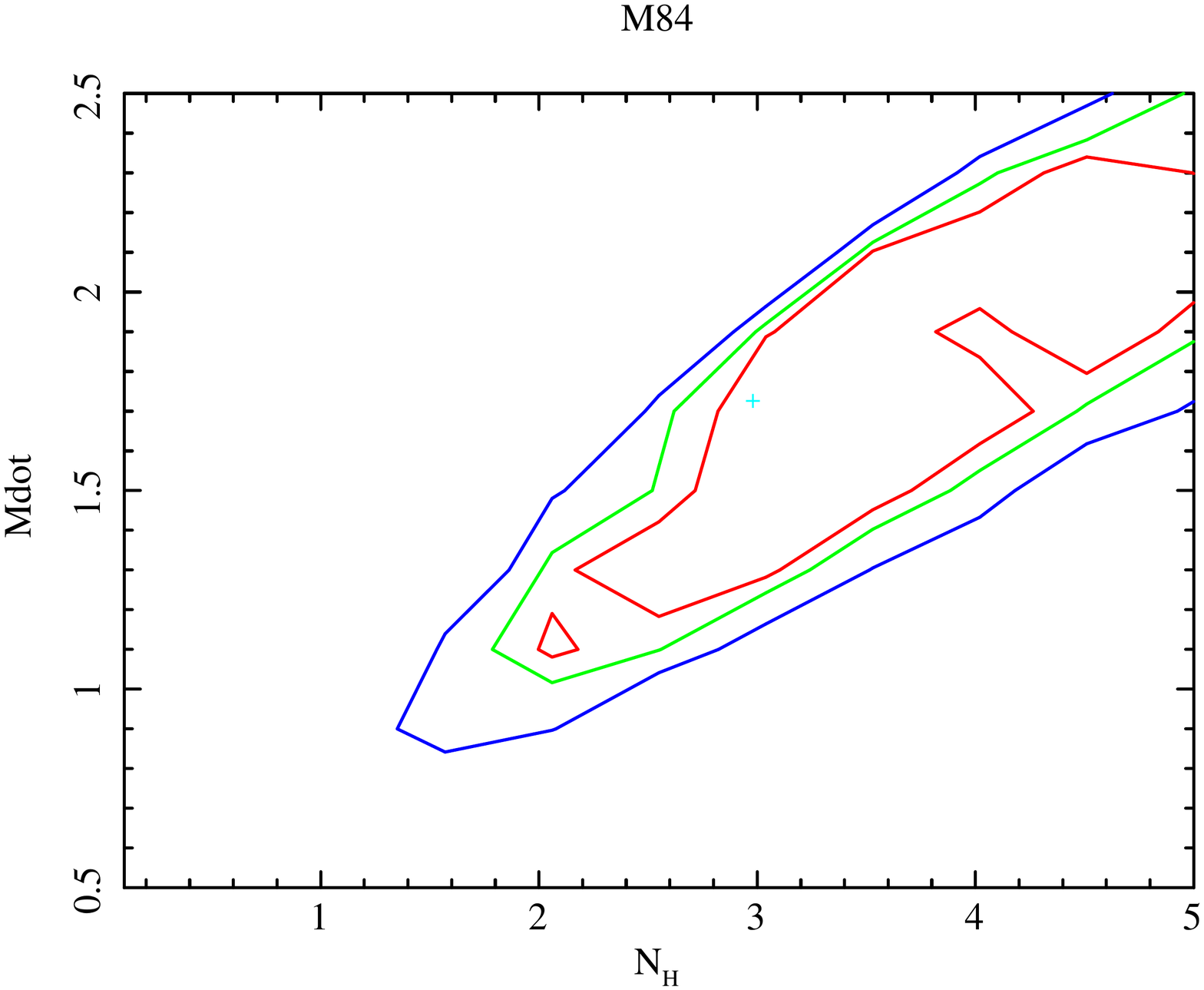}\hspace{-0.85cm}
    \includegraphics[width=0.36\textwidth]{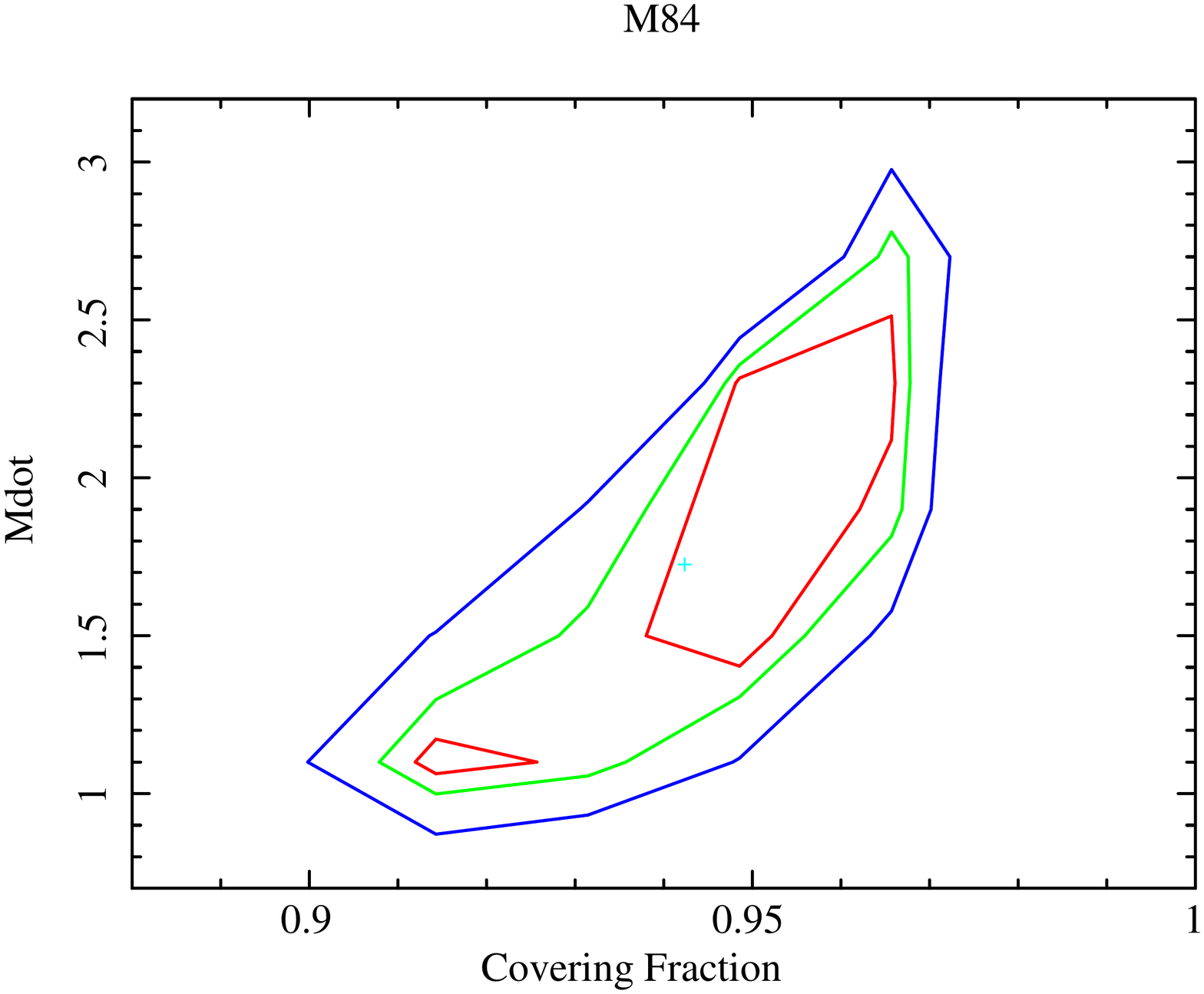}
    \caption{M84, with details as in Fig. 2. }
\end{figure*}
 
 \begin{figure*}
    \centering
    \includegraphics[width=0.36\textwidth]{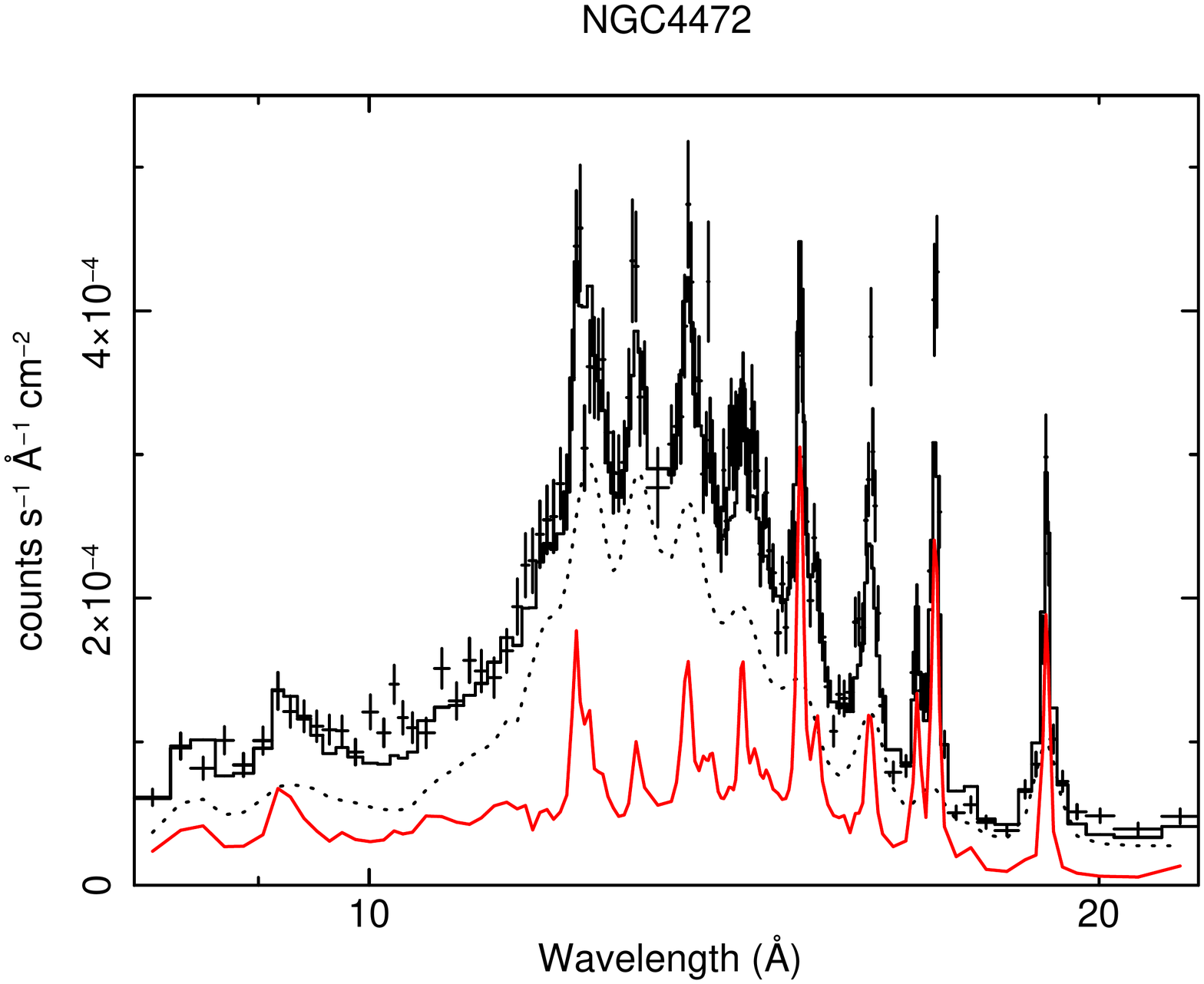}
        \hspace{-0.85cm}\includegraphics[width=0.36\textwidth]{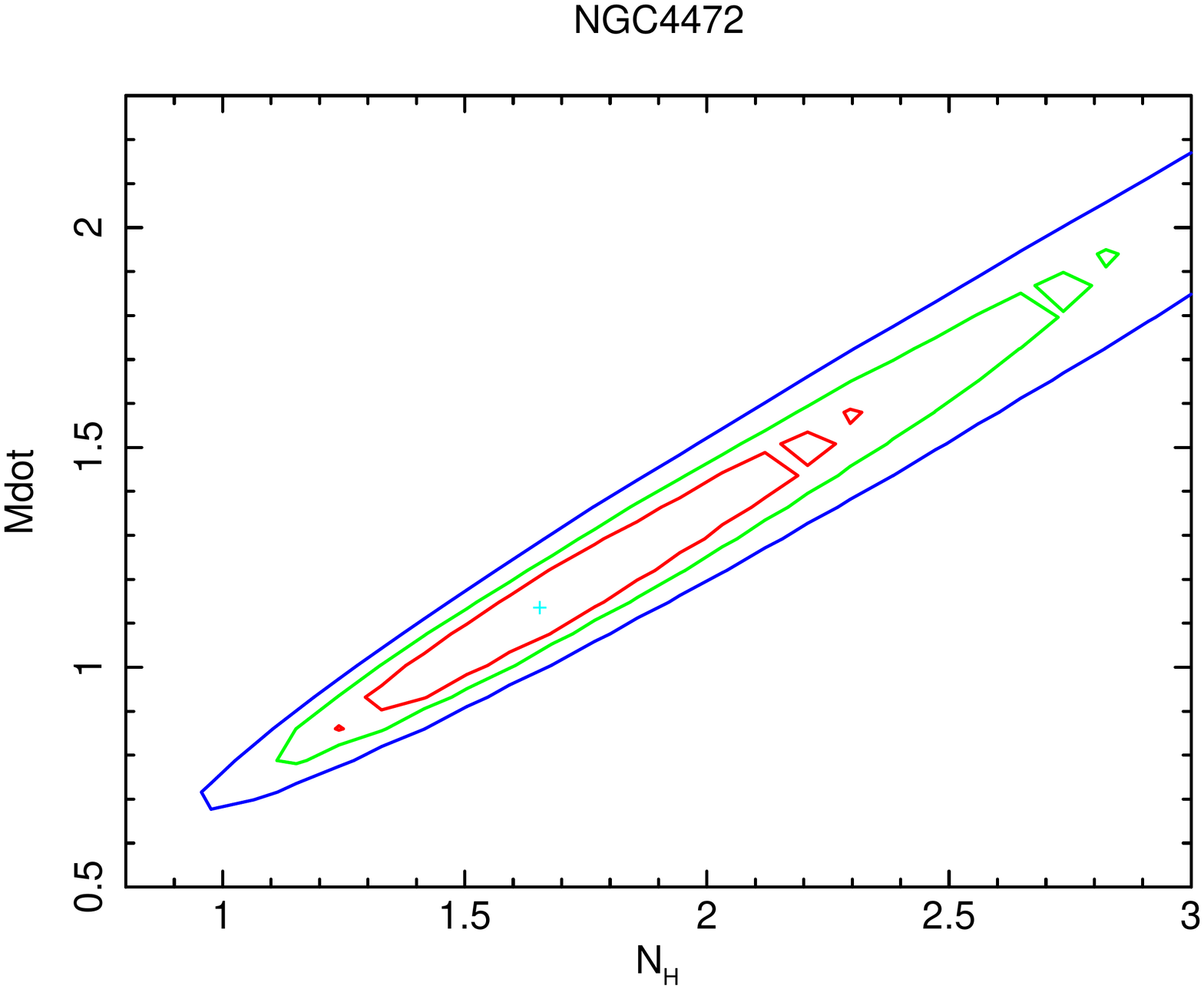}\hspace{-0.85cm}
    \includegraphics[width=0.36\textwidth]{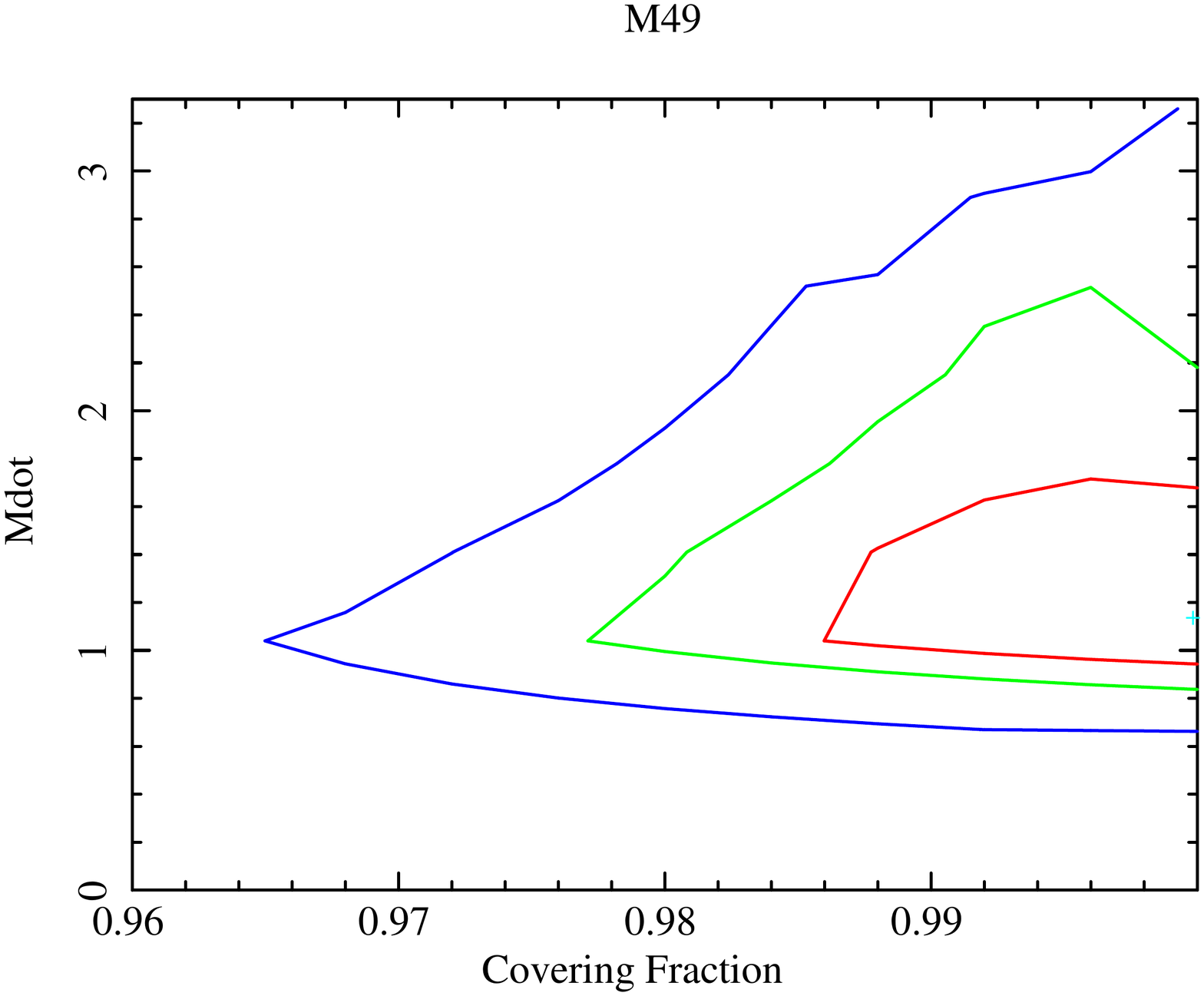}
    \caption{M49, with details as in Fig. 2. }
\end{figure*}

\section{Spectral Analysis}

The datasets were processed using the \emph{XMM-Newton} \textsc{sas} routine \textsc{rgsproc}, using wavelength binning and the extraction regions given in Table 2. 95 per cent of the pulse height distribution was selected to make the spectra. We selected good time intervals from when the rate in bins of 200s was less than $0.3\ps$, for events on CCD 9 with flag values of 8 or 16, outside an absolute cross dispersion angle of $1.5\times10^{-4}$. Background models were created using the regions given for compact sources, or using template background generated by \textsc{rgsbkgmodel} for more extended sources. Spectra from each of the RGS cameras and datasets were combined to make total spectra and responses using \textsc{rgscombine}. 
The spectra were analysed using \textsc{xspec} over the energy range of $8-22$A, which is   where the background is minimised. 

The spectral model used  in HCF1 was {\sc tbabs (gsmooth*apec+gsmooth*mkcflow+gsmooth*mlayerz*mkcflow)}. The model {\sc mlayerz} is explained in HCF1. Noting the similarity to a partial covering model we now use the equivalent model {\sc tbabs(gsmooth*apec+gsmooth(partcov*mlayerz)mkcflow)}. This second version enables us to directly measure the total mass cooling rate of both unabsorbed and absorbed components.   The spectra are shown  (Figs 2--11 left)  together with contour plots of  absorbed mass cooling rate ($\dot M_{\rm a}$) versus intrinsic column density $N_{\rm H}$ (middle) and  covering fraction (right). 

A covering fraction of unity means that all the cooling flow component is absorbed and if zero then none is absorbed. (The Galactic absorption is always present and applied as a frozen parameter.) As the RGS is a slitless spectrometer there is some blurring of the energy scale associated with extended sources. We include that in the model by smoothing the spectral components. The outer {\sc apec} component has
more smoothing $\sigma_6$ than the inner HCF one $\sigma_6^{'}$. When making the contour plots for the less bright objects we often needed to freeze the smoothing parameters to the best fit values in order to have convergence.   
Detailed spectral results are given in Table 2.

\subsection{NGC5044}
Chandra X-ray observations of the Brightest Group Galaxy (BGG) NGC5044 of the X-ray Brightest Group in the Sky, at redshift $z=0.0083$ show a complex cool core with evidence for 2 pairs of bubbles at 2 and 8 kpc along the NW--SE axis \citep{David17, Werner2014}. Extensive optical filamentation and dust absorption are evident (Fig. 12) and cold molecular gas clouds detected. 

The RGS spectral fits clearly reveal an HCF of $20-50 \Msunpyr$ with any unabsorbed flow being less than $0.5 \Msunpyr$ (Fig. 2). The level of FIR emission indicates that the allowed mass cooling rate is nearer to $20\Msunpyr$ (Table 2). 
Absorption was briefly considered for the  lack of gas below 0.6 keV in the early RGS study of \cite{Tamura03}, but  concluded to be "highly unlikely".

Narrow absorption lines are seen against the continuum source in CO observations \citep{David17, Schellenberger20} indicating the presence of small dense clouds.

\subsection{Sersic 159}
The cluster Sersic 159, also known as S1101, at $z=0.0564$, is more X-ray luminous. Chandra imaging of the cool core show that it is disturbed with much of the soft X-ray emission displaced several kpc North of the nucleus \citep{Werner11,McDonald15}. It has dusty optical filamentation stretching out to 40 kpc (Fig. 12). Its RGS spectra were used to demonstrate the lack of the expected cooling flow \citep{Kaastra2001}. 

Here we find a possible HCF flow of up to about $30\Msunpyr$ with any significant unabsorbed flow being less than $10\Msunpyr$ (Fig. 3). The FIR emission would allow up to $30\Msunpyr$ of absorbed flow.  

\subsection{A262}
The low redshift ($z=0.0163$)  cluster A262 shows a complex core in Chandra imaging \citep{Clarke09} with bubbles and a possible tunnel connecting an unusual shaped radio source. 

The RGS spectra (Fig. 4) require an HCF of about $5\Msunpyr$  with no unabsorbed flow. 
The mass and distribution of cold molecular gas in the cluster are discussed  by \cite{Russell2019}.  

\subsection{A2052}
The cluster S2052 lies at $z=0.0353$ and in deep Chandra imaging \citep{Blanton2011} shows 2 pairs of bubbles and surrounding ripples in the core, together with extensive optical filamentation. 

The RGS spectra reveal an HCF of about $10-25\Msunpyr$ with any unabsorbed flow being less than $\sim5\Msunpyr$ (Fig. 5).

\subsection{RXJ0821}
The moderately distant cluster RXJ0821+0752 at $z=0.11$ is very rich in molecular gas ($\sim4\times10^{10}\Msun$) located in a plume several kpc from the nucleus, which hosts a weak radio source. Chandra data also show displaced soft X-ray emission \citep{BayerKim02,Vantyghem19}.

The RGS spectra allow an HCF of $\sim40 \Msunpyr$ with little unabsorbed flow  (Fig. 6).

\subsection{RXJ1532}
RX J1532.9+3021, also known as MACS J1532.8+3021 and hereafter RX1532, is a highly X-ray luminous cluster ($6\times10^{45}\ergps$) at redshift $z=0.36$ \citep{Hlavacek2013}. The cool core has a simple mass cooling rate (from imaging data) of $1000\Msunpyr$, a rate from Chandra spectroscopy of 250\Msunpyr, a central H$\alpha$ nebula of luminosity $3\times10^{42}\ergps$, star formation rate (SFR) of $100^{+50}_{-50}\Msunpyr$ and high molecular gas mass of $8.7\times 10^{10}\Msun$ with a velocity spread in the CO emitting gas of  390\kmps FWHM  \citep[see Fig.3 of][]{Castignani2020}.

The RGS spectra on RXJ1532 (Fig. 7) have a relatively short exposure of 40 ks (Fig. 7). A large HCF of $>1000\Msunpyr$ is allowed together with an unabsorbed flow of $\sim 300-700\Msunpyr$.  

\subsection{MACS  1931.8-2634}
Hereafter MACS1931, it is also a high X-ray luminosity cluster ($4.5\times10^{45}\ergps$) at redshift $z=0.35$ (Ehlert et  al 2011). The cool core has a simple mass cooling rate of $165\Msunpyr$, H$\alpha$ of $2\times 10^{42}\ergps$, SFR of $280\Msunpyr$ and molecular gas mass of $9\times 10^{10}\Msun$ \citep{Castignani2020}.  A velocity spread of over 500\kmps is measured in the molecular gas  \citep[see Fig, 3 in][]{Fogarty2019, Ciocan2021}.

The RGS spectra of MACS1931 (Fig. 8) are also short (45~ks) and allow a large HCF of at least $1000\Msunpyr$ with little  unabsorbed flow.

Both RXJ1532 and MACS1931 show  large cavities/bubbles presumably created by jetted emission from the central AGN, as in many other cool cores.They have spectacular optical line nebulosities and very large molecular gas masses. Reddening is observed across the central parts which indicates distributed dust and  the  likelihood of X-ray absorption adding uncertainty to the measured SFR. $E(B-V)$ maps of the central regions of both BCGs are shown in \cite{Fogarty15}.

\subsection{Phoenix Cluster}
For comparison with the previous 2 clusters, the twice-distant Phoenix cluster at $z=0.596$  has a total X-ray luminosity of  $10^{46}\ergps$, with more extreme H$\alpha$ of $8.5\times 10^{43}\ergps$, SFR of $740\pm160\Msunpyr$ and molecular mass of $2\times 10^{10}\Msun$ \citep{McDonald2019}. It also hosts a powerful central AGN or quasar which accounts for much of its uniqueness. The central AGNs in RX1532 and MACS1931 have detected radio emission but  weak nuclear X-ray emission  detected by  Chandra (the spectrum of the nucleus in M1931 shows intrinsic absorption of $7\times 10^{21}\psqcm$ \citep{Ehlert2011}). \cite{Tozzi2015} determine the absorption towards the QSO nucleus of the Phoenix Cluster to be $46\times 10^{22}\pcmsq$, which means that the AGN emission can be ignored in the RGS band. The RGS spectra of the Phoenix cluster have been studied in detail by \citep{Pinto2018} yielding a mass cooling rate of $350^{+250}_{-200} \Msunpyr$  (without intrinsic absorption) and turbulent velocity of $<300\kmps$ (both measurement uncertainties at 90 percent confidence level).

Here the RGS spectra of the Phoenix cluster allow for a large HCF of  more than $1000\Msunpyr$ (Fig. 9).

\subsection{M84 and M49}
 We end with 2 optically-luminous  elliptical galaxies in the Virgo cluster, M84 and M49, also known as NGC4374 and NGC4472. M84 hosts an AGN which has blown the double-lobed radio source 3C272.1. NGC4472 has a  relatively weak radio source but still has a pair of bubbles. The properties of their cold interstellar media have been studied using Herschel by  \citep{Smith2012,Alighieri13}. The X-ray emission from M84 has been studied with Chandra by \citep{Finoguenov2008} and Bambic et al (submitted) and for M49 by \cite{Biller2004}: see also XMM work by \citep{Su2019}.  Our RGS data indicate  HCFs in both galaxies of $1-2\Msunpyr$ (Figs 10, 11). 

\begin{figure}
    \centering
    \includegraphics[width=0.48\textwidth]{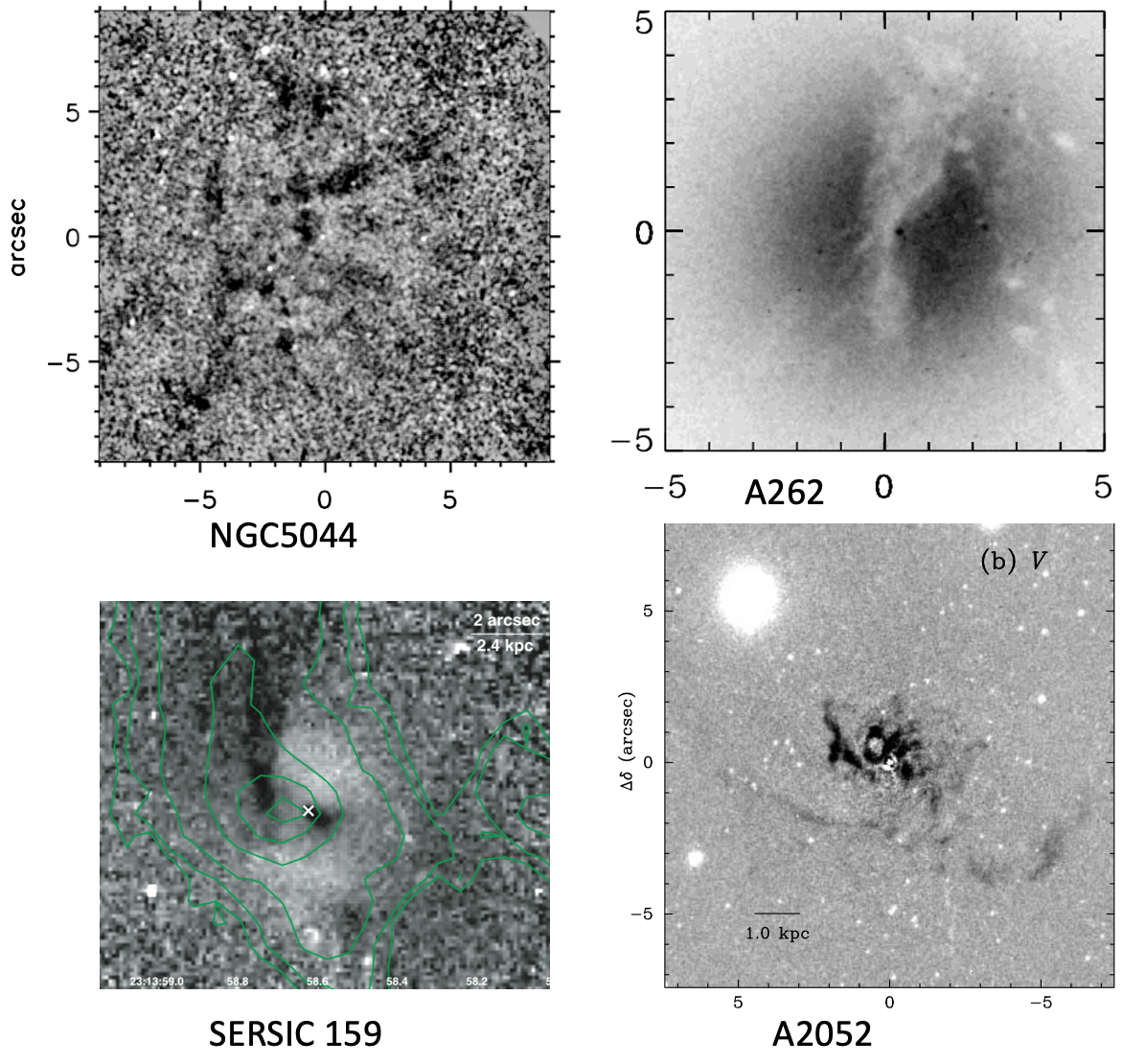}
    \caption{Patchy dust extinction is clearly visible in  published HST images of the central regions of NGC5044 \citep{Temi07}, A262/NGC708 \citep{Wegner12}, Sersic 159 \citep[with radio contours,][]{Werner2013} and A2052/UGC9799/3C317 \citep{Martel02}.   }
    \label{fig:my_label}
\end{figure}

\section{Summary of Results}

All of the 10 objects in this wider sample allow for the presence of an HCF. There is a significant detection of hidden flow in 7 of them. The absorbed luminosity in the cooling flow, calculated from the model spectra using \textsc{xspec}, is generally less than the Far-Infrared luminosity of their dust emission (Table 3), so making the HCF model energetically viable. The tabulated values of $\dot M$ are representative.

Previously the FIR from cool cores has usually been attributed to normal star formation using the \citep{Kennicutt1998} relation,   star formation rate $SFR=L({\rm FIR})/2.32\times 10^{43}\ergps$, and variations on this \citep{Hao2011,Rawle2012}. We do not attempt here to separate the FIR contributions between Hidden Cooling Flows and star formation. Much of $L({\rm H}\alpha)$ also may not be due to star formation but to the absorbed cooling flow radiation field \citep{Polles2021}, so part of that too can be added to $L({\rm FIR})$ when comparing to $L_{\rm a}$ (the total atomic emission is 10-20 times $L({\rm H}\alpha)$ \citep{Ferland2009}).

We have appended the HCF properties of the three clusters in HCF1 (Centaurus, Perseus and A1835) and RXJ1504 (full name RXCJ1504.1-0248) from \citep{Liu2021} to Table 3.

\begin{table*}
	\centering
	\caption{Spectral Fitting Results. The units of column density are $10^{22}\pcmsq$, $kT$ is $\keV$, $Z$ is  abundance  relative to Solar and $\dot M$ is $\Msunpyr$. The smoothing kernels, ${\sigma_6}$ and ${\sigma_6}^{'}$, are defined in $\keV$ at $6\keV$. The redshifts $z$ and some other parameters were fixed (f).}
	\label{tab:properties_table}
	\begin{tabular}{lccccccccccr} 
		\hline
		Cluster & $N_{\rm H}$ & ${\sigma_6}$ & $kT$ & $Z$ &$z$ & $Norm$ & ${\sigma_6}^{'}$ & $CFrac$ & ${N_{\rm H}}^{'}$ &  $\dot M$ & $\chi^2/{\rm dof}$ \\
		\hline
		 & $10^{22}\cmsq$ &$\keV$& $\keV$ & $Z_{\odot}$ & & & $\keV$ & & $10^{22}\cmsq$& $\Msunpyr$\\
		 \hline
		NGC5044 &4.9e-2f & 6.3e-2f & $0.95\pm0.004$ & $0.43\pm0.02$ & 8.3e-3 & 5.9e-2& 1.7e-2 & 0.99 & 2.09 & 21.6 & 1499/1146\\
		Sersic 159 & 1.0e-2f & 3.0e-2f &$2.27\pm0.06$ & $0.31\pm0.02$ & 5.4e-2 & 1.55e-2 & 5.9e-3 & 0.23 & 4.85 & 8.6 & 1336/1341 \\
		 A262 & 7.0e-2f & 5.3e-2f & $1.6\pm0.03$ & $0.31\pm0.02$ & 1.68e-2 & 8.3 e-3 & 5.3e-3 & 1.0 & 0.6 & 6.6 & 1243/1136\\
		 A2052  & 2.85e-2f & 3.22e-2f & $2.07\pm0.07$ & $0.28\pm0.02$ & 3.5e-2 & 1.65e-2 & 1.0e-3 & 1.0 & 0.48 & 13.7 & 1463/1326\\
		 RXJ0821 & 2.1e-2f & 2.2e-2f & $2.1\pm0.3$ & $0.39\pm0.1$ & 0.11 &  1.5e-2 & 1.0e-3 & 0.97 & 1.42 & 60.2 & 595/554\\
		 RXJ1532 & 2.0e-2f & 1.0 e-3f & $4.0^{+4}_{-2}$ & $0.16^{+0.4}_{-0.2}$ & 0.36 & 4.2e-3 & 1.0 e-3 & 0.69 & 0.76 & 1862 & 182/237 \\
		 MACS1931 & 1.0e-2f & 1.0e-2f & $9.1^{+12}_{-3.8}$ & 0.4f & 0.35 & 2.5e-3 & 3.3e-3 & 1.0 & 1.34 & 2175 & 182/186\\
		 Phoenix Cluster  & 1.7e-2f & 5.7e-3f & $7.0\pm0.8$ & 0.5f & 0.597 & 7.4e-3 & 5.0e-3 & 0.70 & 2.5 & 3516 & 947/921   \\ 
		 M84f & 2.9e-2f & 1.7e-2f & $0.82\pm0.01$ & $0.15\pm0.01$ & 3.05e-3 & 8.3e-4 & 1.7 e-2 & 0.94 & 3.0  & 1.73 & 1195/1061\\
		 M49f & 2.0e-2f & 6.3e-2f & $1.0\pm0.01$ & $0.43\pm0.05$ & 3.0e-3 & 1.9e-3 & 1.1e-2 & 1.0 & 1.66 & 1.14 & 801/661\\
	
		\hline
	\end{tabular}
\end{table*}

\begin{table*}
	\centering
	\caption{Relevant Cluster Properties. See subsections in text for individual object references.}
	\label{tab:properties_table}
	\begin{tabular}{lccccr} 
		\hline
		Cluster & $L({\rm FIR})$ & $L_a$ & $\dot M$ & $L({\rm H}\alpha)$ & $M_{\rm CO}$ \\
		\hline
		 & $\ergps$ &$\ergps$& $\Msunpyr$ & $\ergps$ & $\Msun$ \\
		 \hline
		NGC5044 & 3.0e42 & 3.6e42 & 20 &  7.0e40  &1.5e8 \\
		Sersic 159 & 7.3e42 & 5.0e42 & 20  & 2.0e41  & 1.1e9\\
		 A262 & 8.0e42 &1.5e42 & 5 & 9.4e40  &4.0e8\\
		 A2052 & 8.3e42 & 5.8e42 & 20 & 5.8e42 & 2.8e8\\
		 RXJ0821 & 4.5e44 & 7.8e42 & 40 &3.0e41   &3.9e10\\
		 RXJ1532 & 2.3e45 & 2.0e44 &1000 &3e42  &8.7e10\\
		 MACS1931 & 5.6e45 & 4.6e44 &1000 & 2e42 &9.0e10\\ 
		 Phoenix Cluster &3.7e46&3.3e44&2000& 8.5e43& 2e10\\
		 M84 &1.0e42&3.3e41&2.0&4.0e39&<1.8e7\\
		 M49 &1.2e42&2.0e41&1.0&5.8e39&<1.4e7\\
		 \hline
		 Centaurus &3.2e42 &3.6e42&15&1.7e40&1.0e8\\
		 Perseus &5.6e44&5.8e42&50&3.2e42&2.0e10\\
		 A1835 &3.2e45&5.2e43&400&4.4e42&5.0e10\\
		 \hline
		 RXJ1504 &...&1.9e44&520&3.2e43&1.9e10\\
		 \hline
	
		\hline
	\end{tabular}
\end{table*}

\section{Future studies of Hidden Cooling Flows}

More clusters, groups and elliptical galaxies can be explored with new and archival XMM-RGS spectra. The present sample is too small to judge how the properties of the HCF correlate with other parameters but we hope to pursue this in future work. 

We are unaware of any numerical simulations of cool cores that incorporate photoelectric absorption of the cooled gas. It may not be easy but if our work is correct then it is an essential ingredient. It might help identify the scale and strength of turbulence required to mix cooled and cooling gas as well as improve on the simple one parameter intrinsic absorption model. 

The microcalorimeter XRISM  RESOLVE \citep{XRISM2018} is due for launch in 2023. It is similar to the Hitomi calorimeter that was launched and successfully observed the Perseus Cluster briefly in 2016. RESOLVE is expected to have a spectral resolution of about 5eV and 1.3 arcmin spatial resolution and cover a  broader energy band\footnote{The Gate Valve was closed during the Hitomi observation. Excellent spectra were obtained of the iron K complex but no useful spectra were obtained in the HCF region below  2 keV.} than the XMM RGS  enabling  reliable measurements of $\dot M(T)$ over X-ray emitting temperatures. The {\sc xspec} model {\sc mkcflow} we use is a simple constant pressure cooling flow, which may be appropriate over a limited temperature range but not for a flow covering a large radius range. Also, if the gas does flow over a significant radius then gravitational work will be done which can reduce the net inferred mass cooling rate for a given cooling luminosity. For example, if the cooling flow extends from where the gas is at the virial temperature down to the core, the reduction could be a factor of about two. 

Athena X-IFU \citep{Barret2018} spectra obtained at better than 3 eV spectral resolution and 8 arcsec spatial resolution will map the gross distribution of cooled gas and the velocity structure of the various components of the cooling core. We show simulated  XRISM and X-IFU spectra of an arcmin size region at the centre of the Centaurus cluster in Fig. 13. The simulations are based on the {\sc APEC} plus an intrinsically-absorbed and unabsorbed cooling flow spectrum  found in HCF1. The mass cooling rates are $14$ and $0.1\Msunpyr$, respectively.  The exposure time is 100 ks and no background is included in either simulation as our intention is to highlight the bright region between 0.5 and 1 keV which is most sensitive to the HCF. Both spectra can clearly determine the presence and shape of any HCF. The X-IFU can determine the parameters of the HCF, such as total column density, mass cooling rate per unit area,  velocity and velocity dispersion, on 10 arcsec scales. 

AXIS with CCD spectral resolution and 1 arcsec spatial resolution will map cool cores in fine detail. The spatial distribution of mass cooling rate and column density can be correlated with that of the molecular gas and dust to reveal how the hidden cooling flow operates. 

JWST operating in the near and mid infrared will open up cooling cores and enable mapping and quantification of the cold components. It may also reveal whether low mass star formation is occurring at large rates.

\section{Discussion} 
We have found that HCF are common in the cold cores of elliptical galaxies, groups and clusters. We suspect that they are present in all cold cores found in hot galaxy-scale atmospheres. The gas in such cores has been seasoned by stellar mass loss from the host galaxy and is therefore dusty. The energy absorbed from the hot cooling gas is mostly radiated away by the dust. The presence of patchy clouds of dust is the norm for the centres of these objects (Fig. 12).  The absorbed radiation field  leads to low ionization nebulae such as is characteristic of these objects \citep{Polles2021}. 

The fate of the cooled gas is unclear. As mentioned in the Introduction and HCF1, there are several options. Clouds cooled to invisibility is one possibility. As shown in HCF1 the radiative cooling time is short for cold clouds at $5-10\K$ and if they become poor radiators as they cool more then they will not be seen in emission. Many of the AGN in cool core BCGs appear weak or absorbed \citep{HL2011, Rose2022}, which is consistent with a large population of cold, possibly hidden, clouds.

Bubbling AGN feedback can drag cold gas out to larger radii, as revealed by the inverse metal abundance gradients commonly seen \citep{Panagoulia2015, Lakhchaura2019}. Unless that exposes them to heating via AGN feedback, they will, if mixed in at larger radii, only serve to generally increase any cooling flow. Significant gas motions are also associated with cold fronts resulting from past minor subcluster mergers causing the  central potential well to oscillate \citep{ZuHone2016, Fabian2022a}. Further study of the displaced emission in RXJ0821 will be interesting in this respect.

Low-mass star formation promoted by the high pressure environment at the centre of  massive galaxies is plausible due to the reduction in the Jeans mass \citep{Jura1977, Fabian1982, Ferland1994}. As discussed in HCF1, there is evidence for a bottom-heavy IMF in early-type galaxies \citep{vDC2010, Smith2020,Oldham2018}. Evidence for it has also been found in NGC708 in A262 \cite{Wegner12}.  Objects with hidden cooling flows exceeding  $1000\Msunpyr$ (e.g. A1835, RXJ1535, MACS 1931, Phoenix) should then have a very high mass fraction of low mass objects at their centres, with high  central mass-to-light ratios $\Upsilon$. Measurement of the inner mass profile of these galaxies will be an important first step. The pressure in the central kpc of these objects, obtained by extrapolating inward the deprojected density and temperature profiles of their hot gas  \citep[see e.g.][]{Ehlert2011, Hlavacek2013}, is very high $P=nT\sim 10^{7.5}\cmcu\K$. This will further reduce the Jeans Mass to $\sim0.03\Msun$. The physical conditions in the heart of the most massive galaxies in the Universe are not typical of those in lower mass galaxies.

A combination of the above possibilities may be possible. Here we use the $95\%$ aperture spectra which approximately means that we are studying the spectrum from the innermost region with a diameter of about 1 arcmin. The intrinsically-absorbed soft X-ray luminosity is processed and emitted by dusty gas in the FIR, so the angular size of the HCF will be similar to the dust and its FIR emission. This is about one arcmin in Perseus \citep{Mittal2012} and smaller for the other clusters studied here and in HCF1. Absorption in thin dusty filaments further out will contribute, as discussed in HCF1, but only at the level of a few $\Msunpyr$. 

It is likely that HCF vary with time. The cooling time of most of the hot gas within an HCF is significantly less than a Gyr and that of the cooler gas is about 10 million years or less. The objects with the highest HCF rates such as RXJ1504, RXJ1532, MACS 1931 and the Phoenix cluster may be in a shortlived phase of about a Gyr. Such objects are rare. They also contain a central supermassive black hole, which appears inactive in most but is certainly active in the Phoenix Cluster. Quite what that means for the evolution of the extreme HCF remains to be explored. One possibility is that they undergo enormous outbursts exceeding $10^{61}\erg$, such as happened to MS 0735+7421 \citep{McNamara2005, Gitti2007}. 

In a paper that introduced the term "cool core", \cite{Molendi2001} use XMM MOS CCD spectra to investigate whether the gas is multiphase in the cool cores of clusters. Multiphase gas at radii  $r=5-100\kpc$ can lead to an apparent mass cooling rate profile $\dot M\propto r$, as the coolest phases drop out at large radii. Such profiles were inferred from early imaging data. \cite{Molendi2001} found from their spectral analysis that the gas does not show the wide distribution of temperature expected from a multiphase cooling flow. We suggest that any outer cooling flow beyond the regions examined by the RGS is closer to being single phase and that the rising outward temperature profile is established by a balance between AGN heating and radiative cooling. Wide mapping of the temperature structure across an X-ray bright cool core with XRISM Resolve can test this.

\begin{figure}
    \centering
     \includegraphics[width=0.48\textwidth]{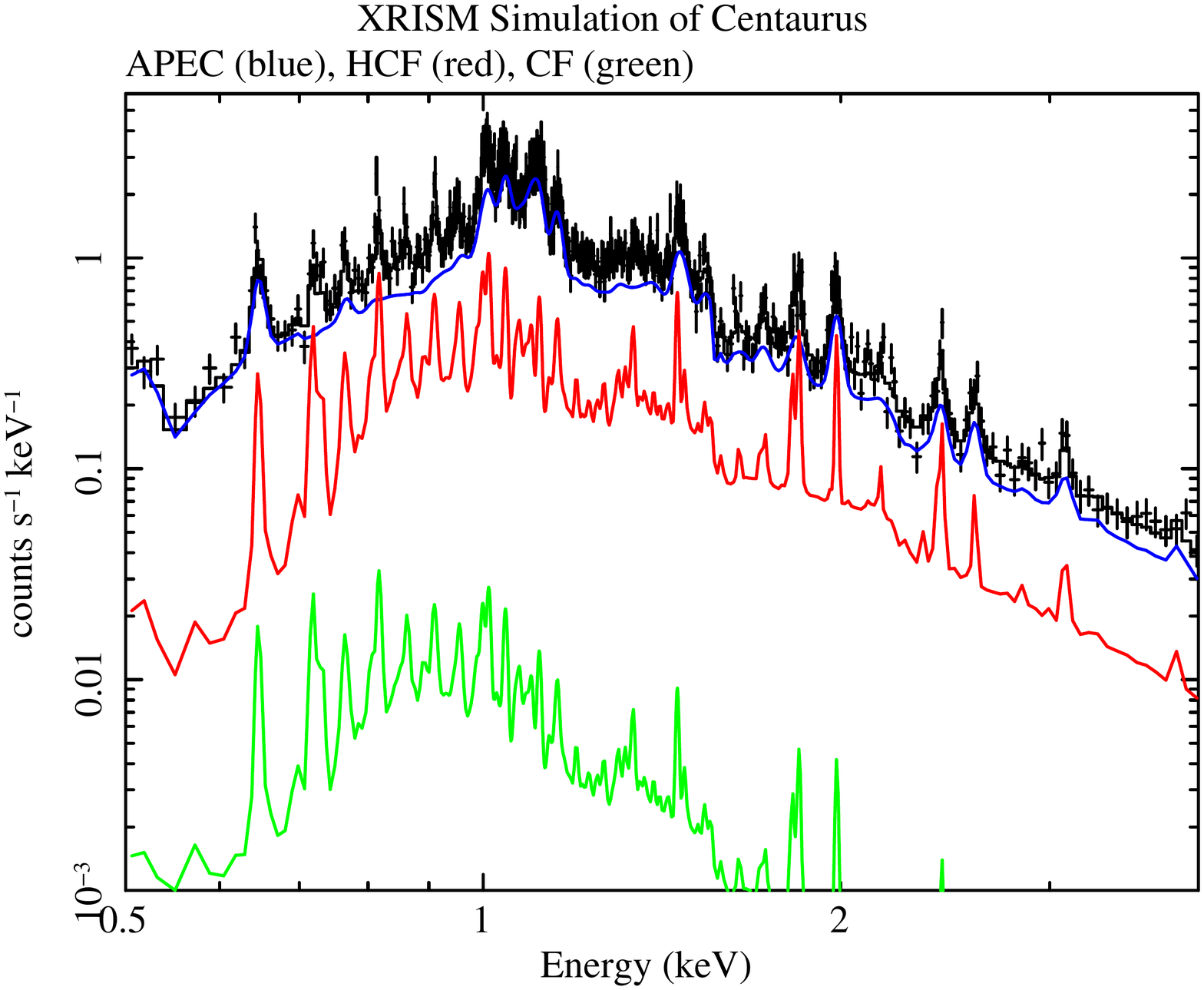}
         \includegraphics[width=0.48\textwidth]{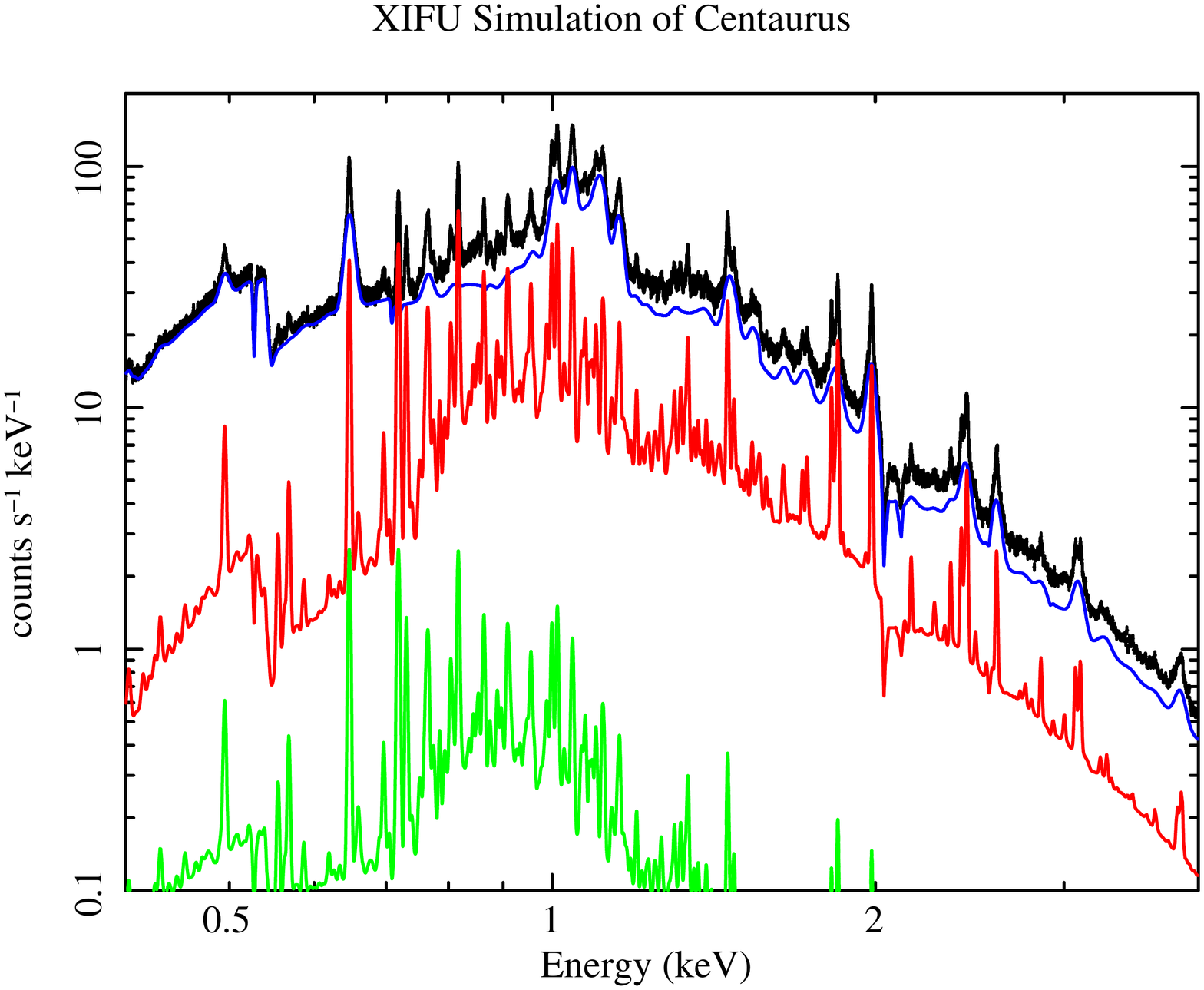}
    \caption{XRISM (top) and XIFU (bottom): Simulated 100 ks spectra of the central arcmin of the Centaurus cluster using the intrinsic absorption model parameters from HCF1, including an intrinsically absorbed flow of $14\Msunpyr$ (red)  and an unabsorbed flow of $0.1\Msunpyr$ (green). The constant temperature APEC component is shown in blue. Note the weaker lines between 0.7 and 1 keV due to the 1.7 keV gas temperature. That energy band contains the FeXVII and other lines indicative of lower temperature gas in the cooling flow.  }  
    \label{fig:my_label}
\end{figure}

\section{Conclusions}
Early RGS analyses were interpreted to claim that there is little or no gas cooling below $T_{\rm V}/3$ in normal clusters of galaxies. The new analysis  with intrinsic absorption presented here can be interpreted as revealing significant levels of gas at temperatures below $T_{\rm V}/3$, consistent with Hidden Cooling Flows of tens to more than $1000\Msunpyr$ in clusters and small Hidden Cooling Flows in elliptical galaxies. Several solutions for the large possible mass of accumulated cold gas implied by  Hidden Cooling Flows have been considered here and in HCF1.

\section{Acknowledgements} ACF again thanks Keith Arnaud for help with defining the models. BRM acknowledges the Natural Sciences and Engineering Research Council for their support.

\section{Data Availability} All data used here are available from ESA's XMM-Newton Science Archive.

\bibliographystyle{mnras}
\bibliography{cool_core_Mdot} % if your bibtex file is called example.bib

% Don't change these lines
\bsp	% typesetting comment
\label{lastpage}
\end{document}